\documentclass[format=acmsmall, review=false]{acmart}
\usepackage{booktabs} 
\usepackage{float}
\restylefloat{table}
\usepackage{dblfloatfix}
\usepackage{placeins}
\usepackage{tikz}

\usepackage[toc,page]{appendix} 
\usepackage{mathtools, cuted}
\usepackage{color}
\usepackage{diagbox}
\usepackage{amsmath}

\usepackage{bbm}

\setcopyright{none}
\renewcommand\footnotetextcopyrightpermission[1]{}
\usepackage[T1]{fontenc}
\usepackage[export]{adjustbox} 

\usepackage{comment}
\usepackage{verbatim,epsfig,times,subfigure, graphicx}
\usepackage{grffile}

\usepackage{url}
\usepackage{hyperref}

\hypersetup{draft}

\usepackage[colorinlistoftodos]{todonotes}

\makeatletter

\providecommand\@dotsep{5}
\renewcommand{\listoftodos}[1][\@todonotes@todolistname]{%
 \@starttoc{tdo}{#1}}
\makeatother

\usepackage[ruled,vlined]{algorithm2e} 

\SetAlFnt{\small}
\SetAlCapFnt{\small}
\SetAlCapNameFnt{\small}
\SetAlCapHSkip{0pt}
\IncMargin{-\parindent}

\setcitestyle{acmnumeric}
\begin{document}
\title[Bandwidth Allocation Games]{Bandwidth Allocation Games}

\author{Niloofar Bayat}
\affiliation{\institution{Columbia University}\country{USA}}
\email{niloofar.bayat@columbia.edu}
\author{Vishal Misra}
\affiliation{\institution{Columbia University}\country{USA}}
\email{vishal.misra@columbia.edu}
\author{Dan Rubenstein}
\affiliation{\institution{Columbia University}\country{USA}}
\email{danielr@columbia.edu}

\begin{abstract}
Internet providers often offer data plans that, for each user's monthly billing cycle, guarantee a fixed amount of data at high rates until a byte threshold is reached, at which point the user's data rate is throttled to a lower rate for the remainder of the cycle. In practice, the thresholds and rates of throttling can appear and may be somewhat arbitrary. In this paper, we evaluate the choice of threshold and rate as an optimization problem (regret minimization) and demonstrate that intuitive formulations of client regret, which preserve desirable fairness properties, lead to optimization problems that have tractably computable solutions.

We begin by exploring the effectiveness of using thresholding mechanisms to modulate overall bandwidth consumption. Next, we separately consider the regret of heterogeneous users who are {\em streamers}, wishing to view content over a finite period of fixed rates, and users who are {\em file downloaders}, desiring a fixed amount of bandwidth per month at their highest obtainable rate. We extend our analysis to a game-theoretic setting where users can choose from a variety of plans that vary the cap on the unbounded-rate data, and demonstrate the convergence of the game. Our model provides a fresh perspective on a fair allocation of resources where the demand is higher than capacity, while focusing on the real-world phenomena of bandwidth throttling practiced by ISPs. We show how the solution to the optimization problem results in allocations that exhibit several desirable fairness properties among the users between whom the capacity must be partitioned.
\end{abstract}


\maketitle


\section{Introduction}\label{sec:intro}
ISPs must limit the rate at which data can be consumed by clients, as there is no unlimited resource in reality. For instance, in 2014, Microsoft unveiled an unlimited data plan for OneDrive, costing \$6.99 per month. A portion of users then started utilizing this plan for extreme data-intensive operations 
and as a result, Microsoft discontinued the plan entirely. Similar issues have been observed by other providers, such as Sprint~\cite{Business} and Comcast ~\cite{Comcast}. While ISPs still claim to offer ``unlimited'' plans, the truth is that their capacity is limited, and hence, they must {\em throttle}~\cite{limiteddata}. Some ISPs may choose to be selective about how they throttle. For instance, they may stop throttling once they notice users are running speed tests~\cite{Throttled}). However, they still must protect their limited bandwidth, even for preferred services (e.g., T-Mobile's "BingeOn" streaming service is throttled~\cite{kakhki2016bingeon}).

To our knowledge, understanding how to throttle with minimal discomfort to the consumer has not been rigorously studied before. In this work, we consider ISPs with bandwidth constraints who must throttle users to remain within those constraints. Our model assumes the most prevailing approach, which is to allow users, on a monthly basis, to download at any rate they can achieve until their total data downloaded for that month reaches a {\em threshold}, at which point they are {\em throttled} to a fixed, lower rate. In practice, users with the same plan will have the same threshold and be throttled down to the same rate. User plan cycle-start-dates are staggered across days of the month to prevent feast periods (when all users download at their individual achievable rates) and famine periods (when all users are throttled).

The choice of threshold and throttling rate can affect a user's regret. Assuming the ISP has a fixed capacity of bandwidth it can support, and user start dates are appropriately staggered, the throttling mechanism can be used to control this aggregate use. This aggregate use will increase with larger thresholds (as a larger percentage of users will download at high rates at any given time), but can be decreased by dropping the post-threshold transmission rates (lessening the aggregate contribution of those actively being throttled). Hence, there is an inherent tradeoff to consider: how far back to push the threshold before rate-limiting users, versus the stringency of the rate limit. To explore this tradeoff, we construct regret functions representing the users' dissatisfaction as a result of being throttled. This dissatisfaction grows with the time over which users are throttled but lessens due to a corresponding increase in the throttled rate. We formally pose the ISPs objective as the minimization of the sum of its users' regrets. We then derive computationally efficient means for calculating the thresholds and rates that achieve the minimum aggregate regret.

We begin by evaluating how effective this thresholding mechanism is when applied to users with staggered starts to their monthly cycles. We show that for sufficiently large numbers of users, the variance in aggregate utilization in throttled settings is roughly equivalent to what occurs in non-throttled settings, but with a lower mean (due to the throttling). In other words, throttling reduces aggregate utilization without introducing "feast" and "famine" sub-intervals within the month.


Our exploration then turns to how an ISP might evaluate which threshold/rate combination best suits its needs as well as that of its users. This problem is formulated as a regret minimization optimization, in which we separately consider the needs of two different types of users: the ``streaming'' user's satisfaction depends on the rate of delivery. Since a reduction of rate affects the quality of the stream, their satisfaction cannot be improved by spreading the transmission over more time. In contrast, the ``file download'' user seeks to download a fixed amount of data at the fastest rate possible, but their satisfaction depends more on the total bytes transferred as opposed to the specific rate at which bytes are transferred: they can adapt to a slower transmission rate by increasing the amount of time they actively download. 


We next explore two types of users within a single-tier system, where all users abide by the same threshold and receive the same rate post-threshold. We establish that for both types of users there are natural classes of regret functions. By formulating an optimization objective as the minimization of aggregate regret, we show that both user types have easily computable thresholds and rates that achieve this minimization.
%
%
We then explore a 2-tier system in a game-theoretic context where the ISP offers two data plans. The users greedily select the tier that minimizes their ``individual'' regret, and the ISP reacts by adjusting each tier's allocation, threshold, and rate to minimize the ``aggregate'' regret. 
Our 2-tier analysis allows exploring the stabilization properties of this game, and how to efficiently compute optimal stabilized offerings. We then show how our approach can be generalized to a larger ($\geq 3$) number of tiers.

Our main contributions are as follows:
\begin{itemize}
    \item We propose a fair allocation of ISP's data resources to heterogeneous users with different demands that may change following a diurnal pattern. 
    
    \item We formulate setting byte thresholds and rate bounds of ISP as a regret minimization problem separately for ``file download'' and ``streaming'' users.
    
    \item We show that for intuitive classes of regret functions, the optimal choice of threshold and rate is efficiently computable, and satisfies a set of desirable fairness properties (with respect to the relative regrets in non-optimal choices).
    
    \item We generalize our model in the context of ``file download'' users and consider settings where each user can choose from different ISP plans. We show how the optimization translates to a mechanism resembling a Stackelberg game, and show how to efficiently compute the ISP's threshold-rate pairs for different plans.

\end{itemize}

The rest of the paper is organized as follows. \S\ref{sec:model} describes the model for users and the ISP, \S\ref{sec:reg_model} describes our regret minimization model separately for streaming (Algorithm \ref{alg:reg_min_stream}) and file downloads (Algorithm \ref{alg:reg_min}). \S\ref{sec:multiple_tiers} generalizes the model when ISP has more than one tier through a game theoretic approach (Equation \ref{eq:2tier_reg}). \S\ref{sec:related} presents a literature review, and finally the paper is concluded in \S\ref{sec:conclusion}

\section{Model} \label{sec:model}

Our model includes a description of the user, the ISP, and the throttling mechanism. Table \ref{Tab:parameters} summarizes the parameters of this paper. Let $\mathcal{N} = \{1,2,\ldots, N\}$ be a set of users. When user $i$ is actively transmitting or receiving, it has a {\em desired rate of transmission}, $R_i$, \footnote{Note that we do not differentiate between upload and download, and for simplicity, assume that a single rate applies for both directions of transfer. We use both terms of ``transmit'' and ``receive'' to address users' transfers. Our results are easily extendable to the case where these rates differ.} where $\boldsymbol{R}$ is the set of user rates, $\{R_i\}$. Without loss of generality, we assume that the users are numbered such that $R_i \le R_j$ for $i<j$.

Not all users actively transmit at all times. Let $0 < x_i \le 1$ represent the fraction of time user $i$ transmits, such that its transmission rate in an unthrottled system would average to $x_i R_i$. If $R_i$ represents bits per second, user $i$'s average bits sent in a (billing) cycle would be $x_i R_i \mathcal{T}$ where $\mathcal{T}$ is the number of seconds per cycle. for simplicity, we specify $R_i$ as \emph{bits per cycle} so we can drop the $\mathcal{T}$ from the equation. 

We assume an ISP allocates a fixed data capacity, $\mathcal{C}$, per (monthly) cycle that it can apportion among its users. When $\sum_i x_i R_i > \mathcal{C}$, user demand exceeds ISP capacity, and the ISP must do something to reduce overall user consumption. The conventional approach is to set a threshold $T$ and rate $r$ (also specified in bits/cycle) such that a user $i$'s transmission can proceed at rate $R_i$ until its use (in bits) equals $T$. Thereafter, for the remainder of the cycle, its transmissions are restricted to rate $r$. If all users started their cycles at the same time, the ISP would experience a heavy initial (possibly overwhelming) deluge, followed by a much lighter demand at the end of the cycle. As we show in \S\ref{sec:stagger}, this phenomenon is prevented by staggering the start times of the various users uniformly across the interval of a cycle (i.e., on different days of the month). This way, the demand at any instant in time consists of roughly the same ratio of rate-unconstrained and rate-constrained users such that this threshold-rate approach generally flattens peak demand.

Note that using the units as defined above, user $i$ only reaches the threshold when $x_i R_i \geq T$, and when so, the threshold is reached at the fraction of time $t_i = T / (x_i R_i)$ after user $i$'s start date within the cycle. For the remaining fraction $1 - t_i$ of the cycle, the user, when active, can only receive at the rate $r$. Furthermore, the ISP may offer different plans with different $T$ and $r$. For simplicity of presentation, we defer addressing multiple plans to \S\ref{sec:multiple_tiers}.

How a user reacts to a reduced rate depends on whether the user is streaming or (file) downloading. A streamer experiences a degraded quality of its actively watched stream and cannot improve the quality by increasing the fraction of time it is active. Hence after reaching the threshold, a streamer would continue to download at the reduced rate for the same fraction $y_i = x_i$ of the time it used prior to the rate reduction. The number of bits consumed by a user $i$ who reaches the threshold and gets throttled within a cycle is therefore $x_i R_i t_i + y_i r (1-t_i)$ with $t_i = T / (x_i R_i)$, reducing to $T + y_i r (1 - t_i)$. In contrast, a downloader $i$ would seek to download a total of $x_i R_i$ bits within a cycle. When the ISP caps the transmission rate, the user could increase the amount of time spent downloading. Hence, the user can (and we assume would) adjust the fraction of time it is active to $y_i \ge x_i$ after being rate throttled in an attempt to maintain a total download of $x_i R_i$ per cycle.

\begin{table}[t]
\centering
\footnotesize
\begin{tabular}{c c} 
\toprule
parameter & description \\
\hline
$\mathcal{N}, N$ & the set of ISP users and its size, respectively\\
\hline
$\mathcal{H}$, $H$ & the set of throttled users and its size, respectively \\
\hline
$\mathcal{L}$, $L$ & the set of unthrottled users and its size, respectively \\
\hline
$\mathcal{L}_D, \mathcal{L}_R$ & low-demand users, low-rate users ($\mathcal{L} = \mathcal{L}_D\cup \mathcal{L}_R$) \\
\hline
$R_i,\boldsymbol{R}$ & pre-throttling rate of user $i$, vector of user rates\\
\hline
$B_i$ & bandwidth allocation of user $i$ \\
\hline
$T,r$ & throttling threshold, and throttling rate\\
\hline
$x_i,\boldsymbol{X}$ & the fraction of time user $i$ utilizing the data pre-throttling, $x_i$ vector for all users \\
\hline
$y_i,\boldsymbol{Y}$ & the fraction of time user $i$ utilizing the data post-throttling, $y_i$ vector for all users \\
\hline
$\mathcal{C}$ & the network capacity \\
\hline
$\mathfrak{R}$, $\mathfrak{R}_i$, $\mathfrak{R}^*$ & aggregate regret, user $i$'s regret, minimum regret \\
\hline
$T^*$ & $T$ minimizing the regret \\
\hline
$\hat{T}$ & maximum $T$ when $r = 0$ \\
\hline
$\rho, \tau$ & the exponents of regret function for rate-term, time-term \\
\hline
$p_j$ & the normalized price of tier $j$ in a multi-tier ISP \\
\hline
$\boldsymbol{V}$ & the set of available video codecs \\
\hline
$\kappa$ & tier price coefficient \\
\bottomrule 
\end{tabular}
\caption{Summary description of parameters.}
\label{Tab:parameters}
 \end{table}

\subsection{Constraints on $T$ and $r$}\label{sec:const-T-r}

For a given $T$ and $r$, not all users are necessarily throttled, as some users' monthly consumption may lie below the threshold $T$ (low-demand users); others may reach the threshold but using a rate $R_i$ that is smaller than the throttled rate $r$ (low-rate users). For a given $T$ and $r$, let $\mathcal{H}(T,r) \subseteq \mathcal{N}$ be the set of users who are throttled and $\mathcal{L}(T,r)$ be the set of users who are not throttled, with $\mathcal{L}(T,r)$ being further partitioned into low-demand users ($\mathcal{L}_D(T,r)$), and low-rate users ($\mathcal{L}_R(T,r)$). The ISP must select $T$ and $r$ such that its capacity is not exceeded. Noting that having the capacity greater than the total consumption would unnecessarily constrain user access, the ISP's goal is to achieve the equality:

\begin{equation}
\label{eq:uni_cap} 
\mathcal{C} = \sum_{i\in\mathcal{L}(T,r)} R_i x_i + \sum_{i\in \mathcal{H}(T,r)}(T + r y_i (1-\frac{T}{x_i R_i})). 
\end{equation}

\begin{figure}[t]
\centering
\subfigure[Before throttling]{\includegraphics[width=0.3\linewidth, angle=0]{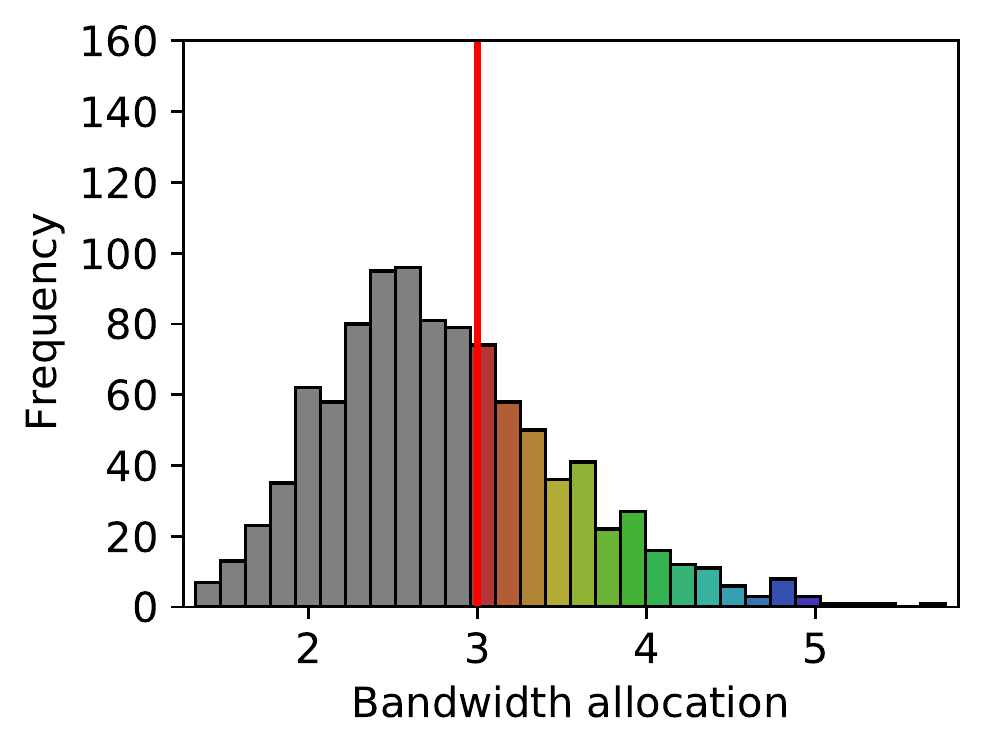}} 
\subfigure[After throttling]{\includegraphics[width=0.3\linewidth, angle=0]{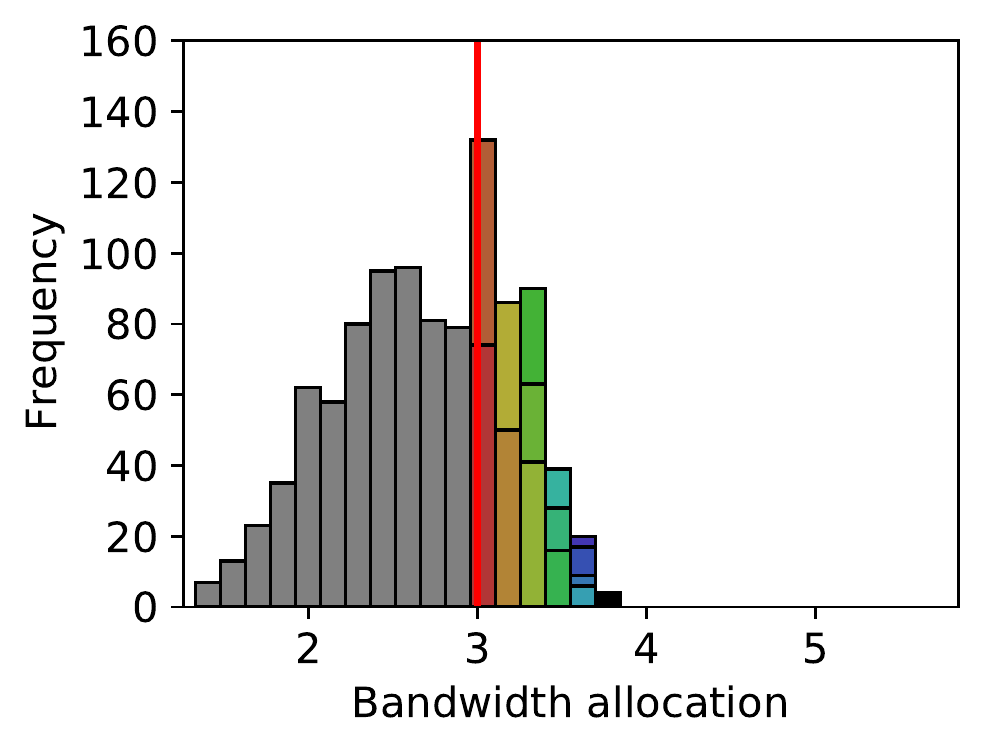}} 
\caption{Bandwidth allocation histogram for $1000$ users drawn from a log-normal distribution with $\mu =1, \sigma = 0.25 $, before and after throttling. The users total bandwidth requirement is $\approx 2800$ (a). When throttling, we have $T=3.0$, $r = 1.9$, and $\mathcal{C}=2700= total\ bandwidth\ allocation$ (b).}
\label{fig:new_discrete_1}
\end{figure}

For ease of presentation, from here on, we drop the $T$ and $r$ when specifying $\mathcal{H}(T,r), \mathcal{L}(T,r), \mathcal{L}_D(T,r)$ and $\mathcal{L}_R(T,r)$. To visualize the effect of $T$ and $r$ on client consumption, consider $1000$ users whose rates $R_i$ are drawn from a log-normal distribution \cite{antoniou2002log} with parameters $\mu =1$ and $\sigma = 0.25$. Figure \ref{fig:new_discrete_1}a depicts the distribution of user rates as a histogram, which also maps to desired consumption within the cycle. If we assume that $x_i = 1$ for all users, then aggregate consumption exceeds $2800 GB$. What happens if $\mathcal{C} < 2800$ such that thresholding is required? A revised distribution of consumption per user is shown in Figure \ref{fig:new_discrete_1}b when $\mathcal{C}=2700$ and after throttling is applied. We see that users with low rates (below the threshold of $3$ as indicated by the vertical red bar) are unaffected, whereas high-rate users' consumption mostly drops.


\begin{figure}[t]
\centering            
\includegraphics[width=0.4\linewidth, angle=0]{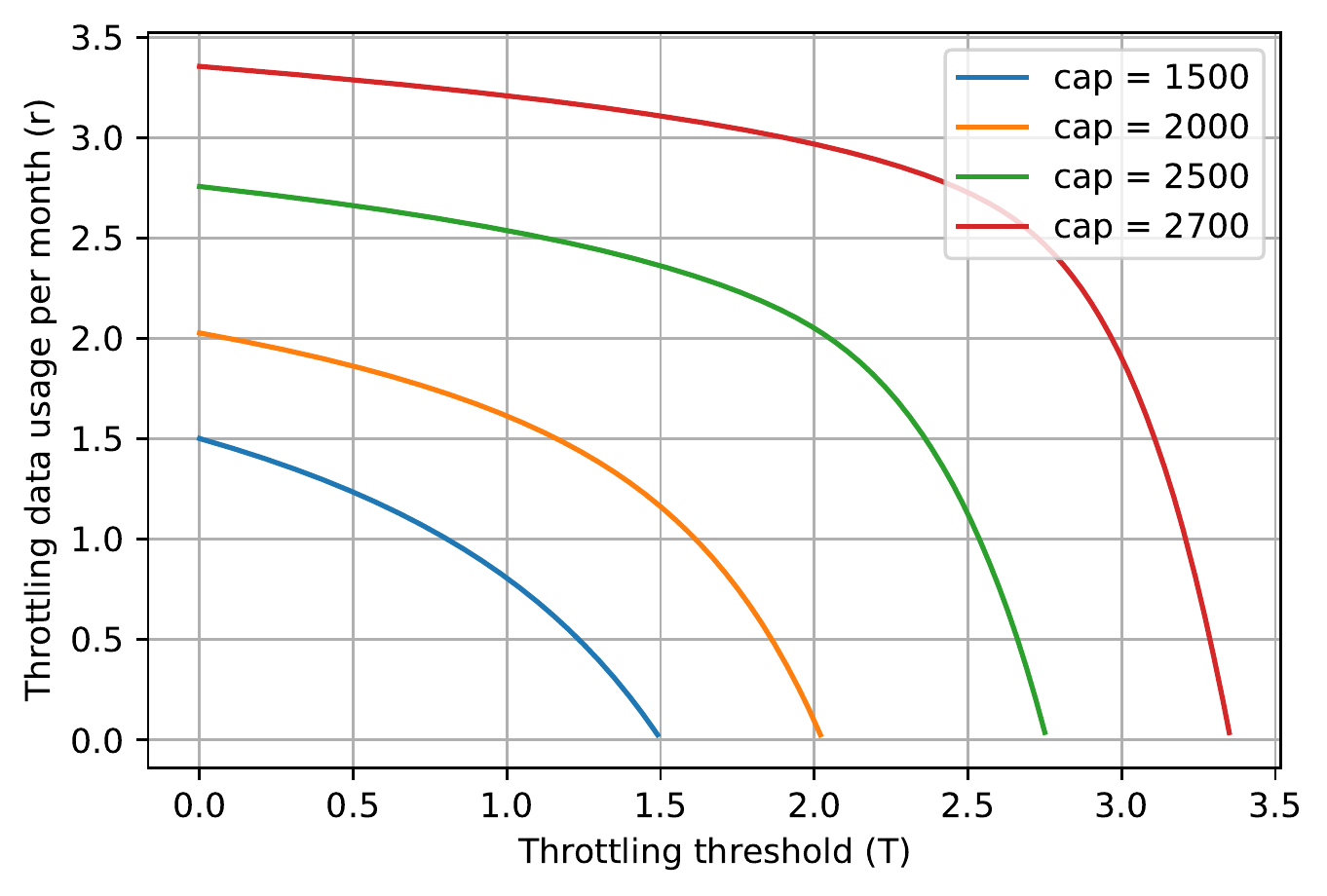}{} 
\caption{$T$ versus $r$ for different ISP capacities where $1000$ users drawn from a log-normal distribution with $\mu =1, \sigma = 0.25 $, where the users total bandwidth requirement is $\approx 2800$.
}
\label{fig:discrete_rT1}
\end{figure}

Figure \ref{fig:discrete_rT1} depicts the inherent tradeoff of choosing a threshold $T$ versus a corresponding throttled rate $r$, with the different curves representing different values of $\mathcal{C}$. A small choice of $T$ means that users will quickly reach the threshold, leaving significant capacity to be consumed post-threshold. Hence, a larger throttled rate ($r$) is permissible. As the threshold is increased, more of the capacity will be consumed pre-threshold, forcing a sharper decline in post-throttled rate $r$. An increase in overall capacity $\mathcal{C}$ permits an increased throttled rate for a given threshold.

 \subsection{Staggered Monthly Billing Cycles}
 \label{sec:stagger}
 
As mentioned in \S\ref{sec:intro}, user plans are staggered across the month to prevent feast and famine periods. To best emulate reality via a simple model, we choose user start dates uniformly at random from $30$ days of a month, and their cycles are periodically repeated every $30$ days. A user's rate at any time is described as a 3-state system that can change hourly. With probability $1-x_{i}$, user $i$ is inactive. With probability $x_{i}$, the user is either throttled or non-throttled; if user $i$ utilizes more than $T$ since their start date, they are throttled to rate $r$ for the rest of their cycle, otherwise, their rate remains at $R_i$.

\begin{figure}[t]
\centering
\subfigure[Two non-throttled users]{\includegraphics[width=0.325\linewidth, angle=0]{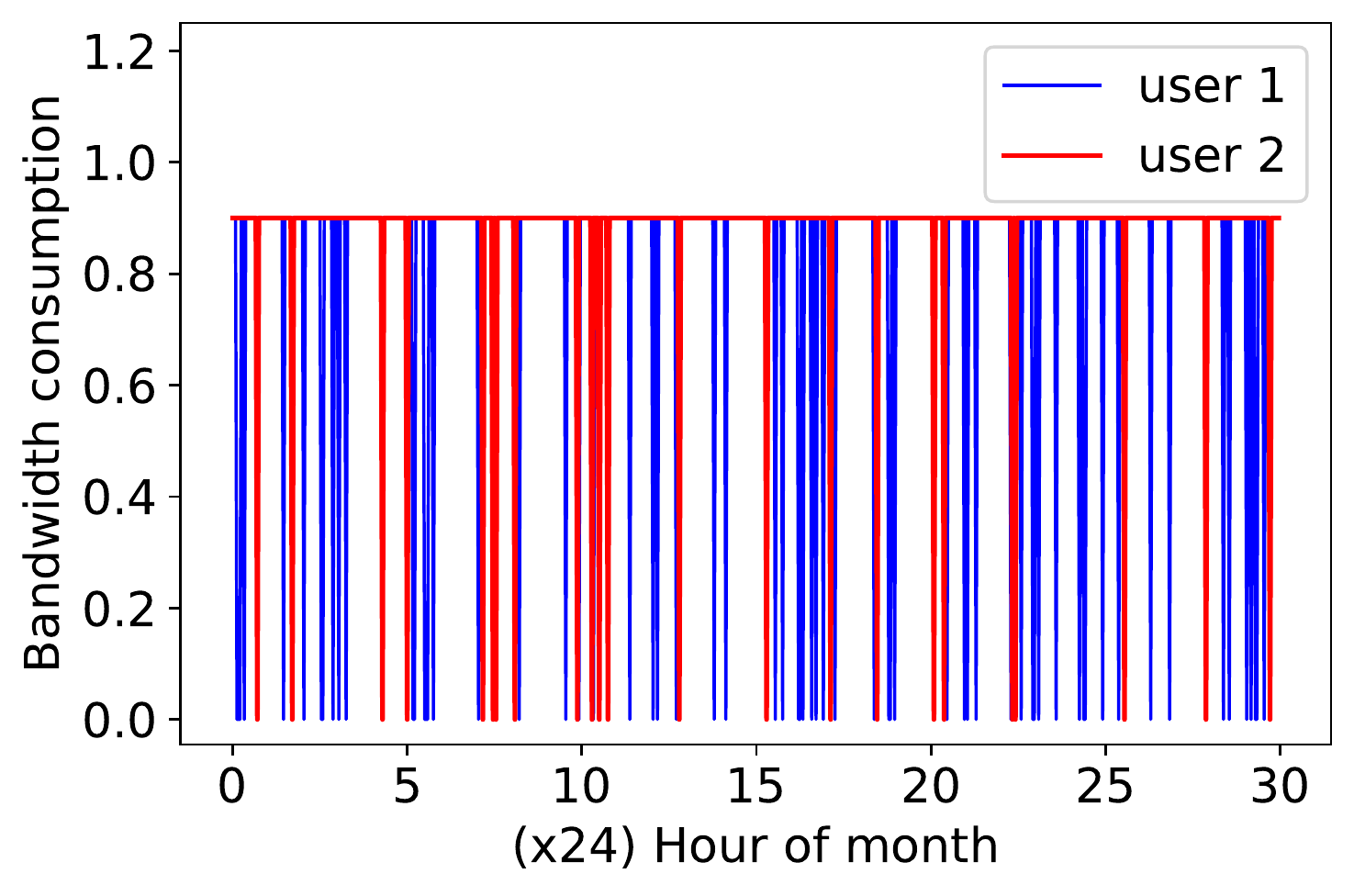}} 
\subfigure[Two throttled users]{\includegraphics[width=0.325\linewidth, angle=0]{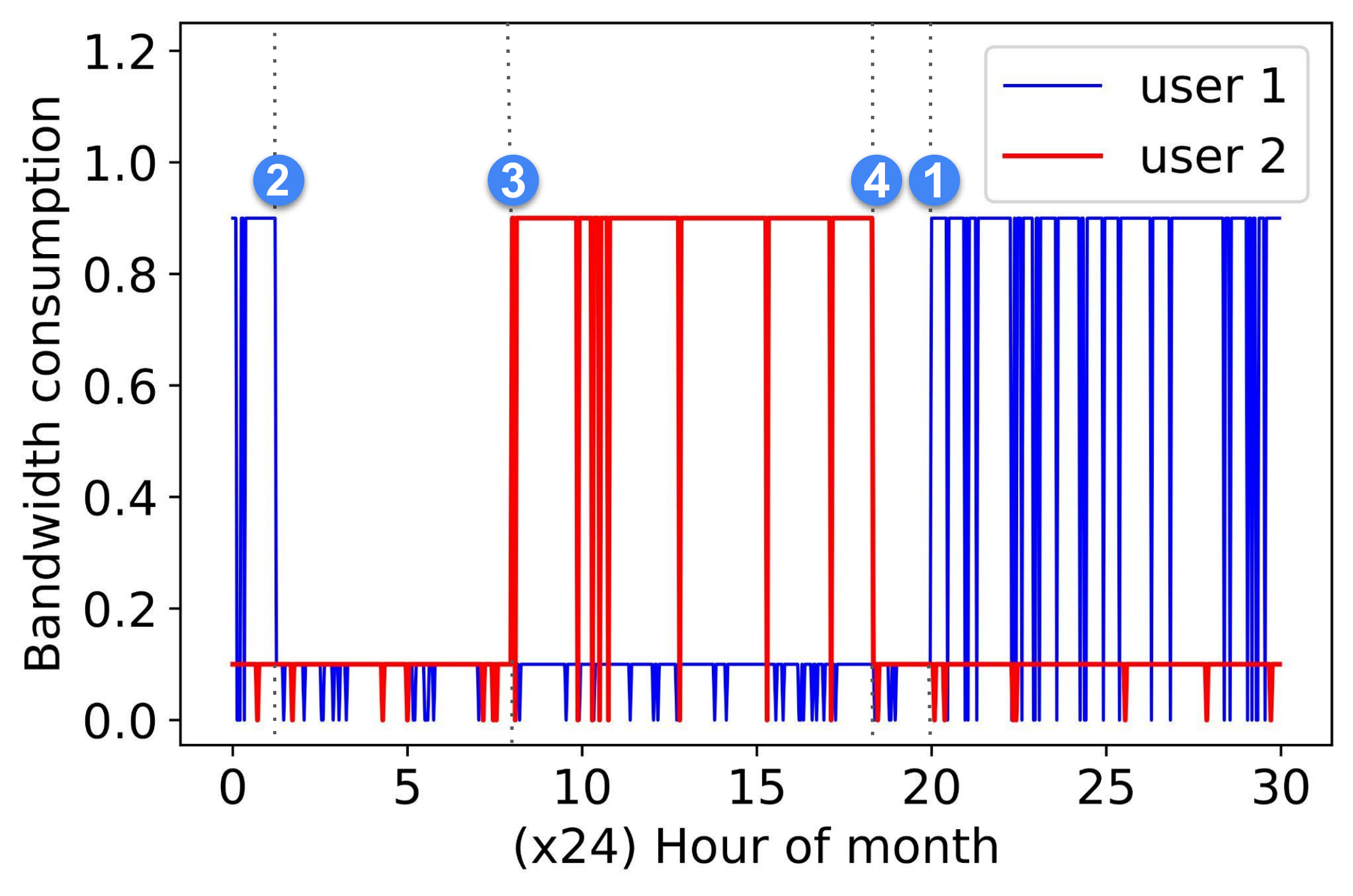}} 
\subfigure[Cumulative 1000 users]{\includegraphics[width=0.325\linewidth, angle=0]{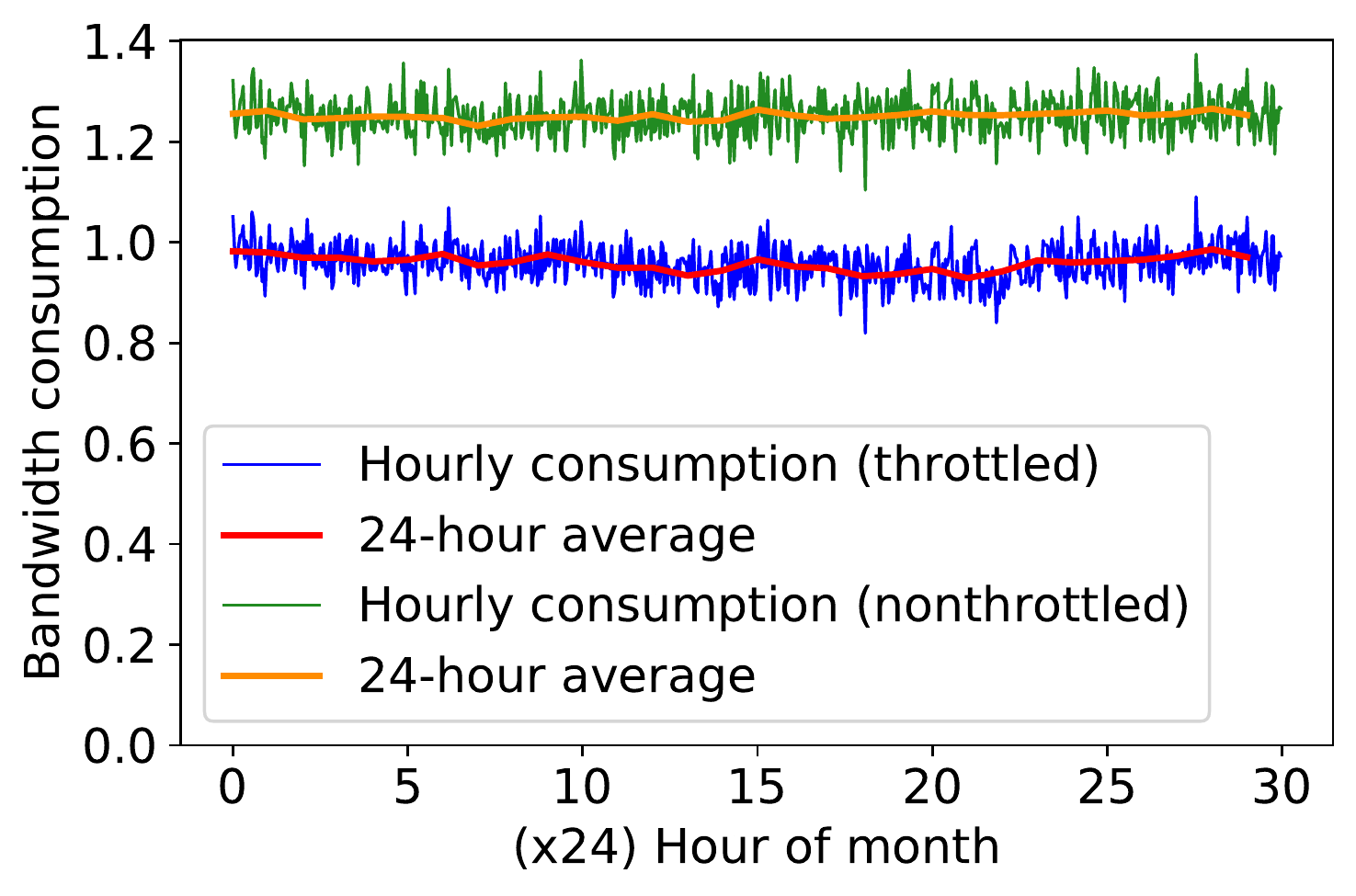}} 
\caption{Users' bandwidth allocation with different $30$-day cycles, where 
$[x_1, x_2]= [.9, .97], [R_1, R_2] = [.9, .9] $. (a) the network can support their demand and they are not throttled, (b) the network capacity is $80\%$ of the demand and they get throttled with $T=.3$, $r = .1$. (c) Cumulative bandwidth allocation and its daily average for $1000$ users, where network capacity is $80\%$ of the demand and $T=.3$, $r = .1$. Each user has three states: non-throttled, throttled, and off. The hourly bandwidth consumption in a\&b is normalized to $1/(24\times 30)$ and in c to $\mathcal{C}/(24\times 30)$. }
\label{fig:ind_user}
\end{figure}

 Figure \ref{fig:ind_user}(a) shows these states for two arbitrary users, where the network capacity is higher than the total demand, and the users do not get throttled. Therefore, users $1$ and $2$ only have two states, non-throttled and inactive. Figure \ref{fig:ind_user}(b), however, shows the same users, where the network capacity is $80\%$ of their demand, $T = .3$ and $r = .1$.  Note that in Figure \ref{fig:ind_user}(a,b),  the y-axis is normalized to $1/(24\times 30)$. Hence, when the y-axis value is equal to $0$, users are inactive, when it is equal to $0.9$, they are non-throttled, and when it equal to $0.1$, they are throttled. Vertical lines marked by numbers $1$, $2$, $3$, and $4$ indicate user $1$'s start of the cycle, user $1$'s start of throttling, user $2$'s start of the cycle, and user $2$'s start of throttling, respectively. Each user gets throttled after consuming $T = .3$, and the duration of throttling is longer for user $2$ (from line 2 to line 1) compared to user $1$ (from line 4 to line 3 of next period), which is because $R_2x_2> R_1x_1$, where $ [R_1, R_2] = [.9, .9] , [x_1, x_2]= [.9, .97]$. 

Finally, if we increase the number of users and compute the cumulative bandwidth consumption, the states of users cancel out each other due to their heterogeneity, which we can see in Figure \ref{fig:ind_user}(c). It shows the cumulative bandwidth consumption of $1000$ users both when the ISP capacity can support the demand (green curve) and when it is $80\%$ of the demand (blue curve). Note that the y-axis in Figure \ref{fig:ind_user}(c), which considers all users in the system, is normalized to $\mathcal{C}/(24\times 30)$. We observe that given a heterogeneous market of users and when users' start dates are distributed throughout the month, famine/feast periods are avoided even if the ISP practices throttling.

In reality, user consumption often exhibits a diurnal behavior \cite{bayat2021down,balachandran2013analyzing}, where rates of utilization vary with time of day. We show that our results hold and famine/feast periods are avoided in daily average usage even when diurnal patterns exist. Assuming users belong to the same timezone, we model the likelihood of user activity as a sine wave. For the back of envelope calculations, we assume the lowest traffic rate occurs between $2 AM-6 AM$ \cite{wang2021examination}. Since on average, user $i$ is active $x_i$ fraction of their cycle, our since wave has an average of $x_i$, period of $24$ hours, and min occurring approximately at $4 AM$, hence the median happening at $10 AM$. Let $x_{ij}$ be the random variable of user $i$ being active at time slot $j$. We must have $0 \leq x_{ij}\leq 1$. Hence:
\begin{equation}
x_{ij} =  0.5 * min(x_i, 1-x_i). sin(\frac{2\pi}{24} (j-10)) + x_i     
\end{equation}
where adding the term $x_i$ ensures the average of $x_{ij}$ equals $x_i$, the multiplicand $min(x, 1-x)$ ensures $0 \leq x_{ij}\leq 1$, and the multiplicand $0.5$ is an arbitrary scale of sine wave, which also avoids all users being active or inactive at a time. The time slot $j$ is the hour of the day in a 24-hour format.

\begin{figure}[t]
\centering
\subfigure[hourly bandwidth consumption and daily average for $100$ users]{\includegraphics[width=0.325\linewidth, angle=0]{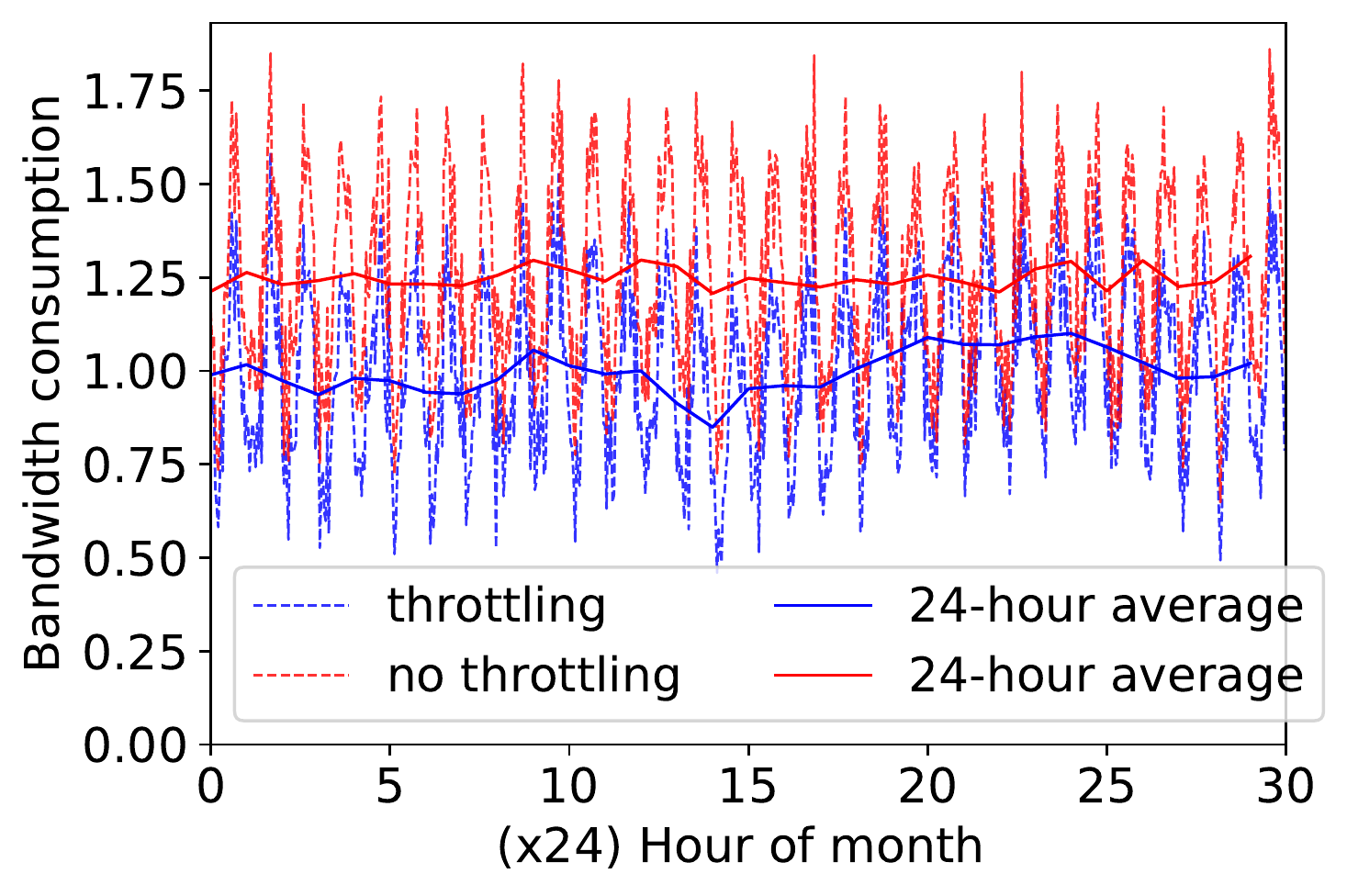}}
\subfigure[hourly bandwidth consumption and daily average for $10,000$ users]{\includegraphics[width=0.325\linewidth, angle=0]{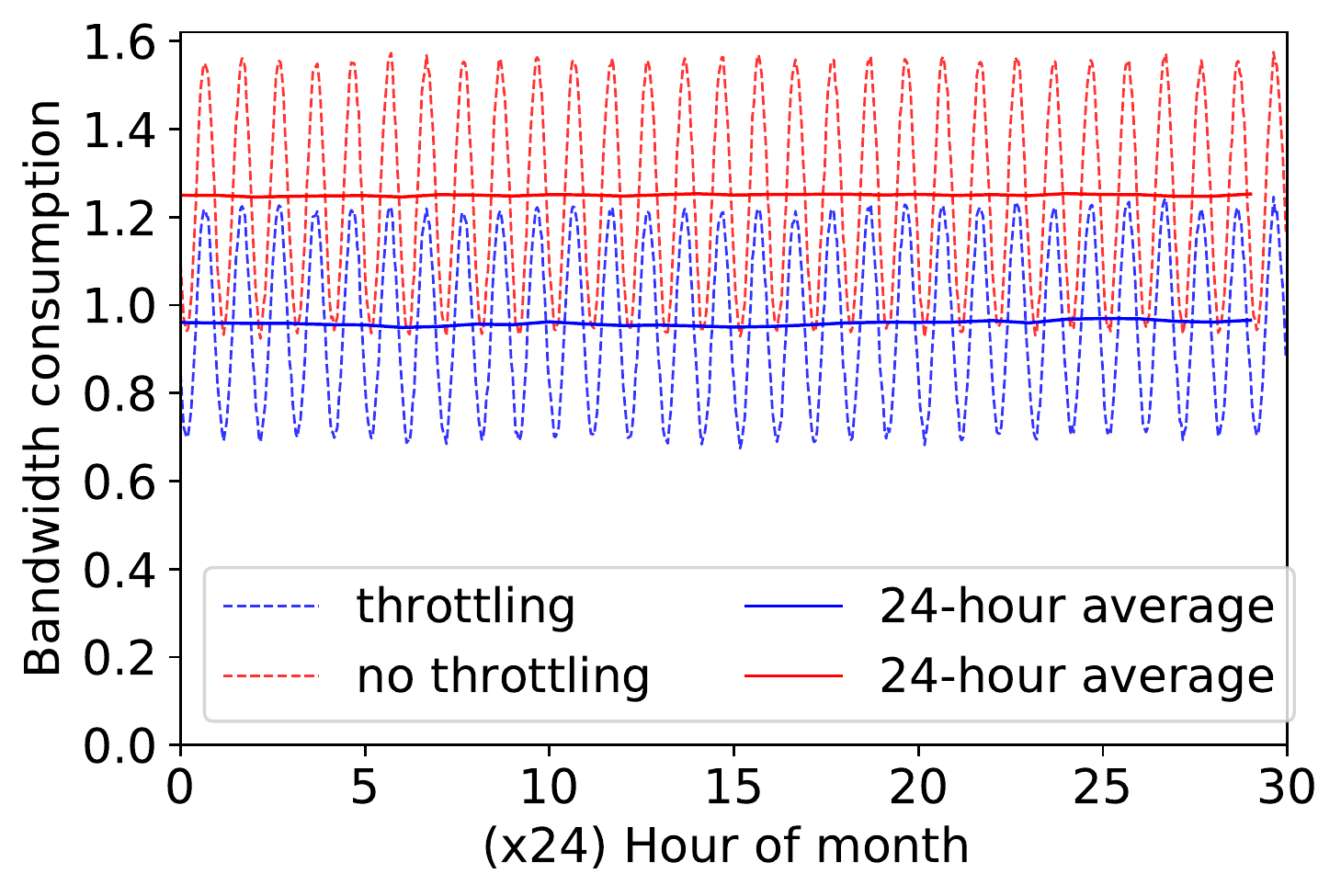}} 
\subfigure[std of the ratio ``throttling/no throttling'' bandwidth consumption]{\includegraphics[width=0.325\linewidth, angle=0]{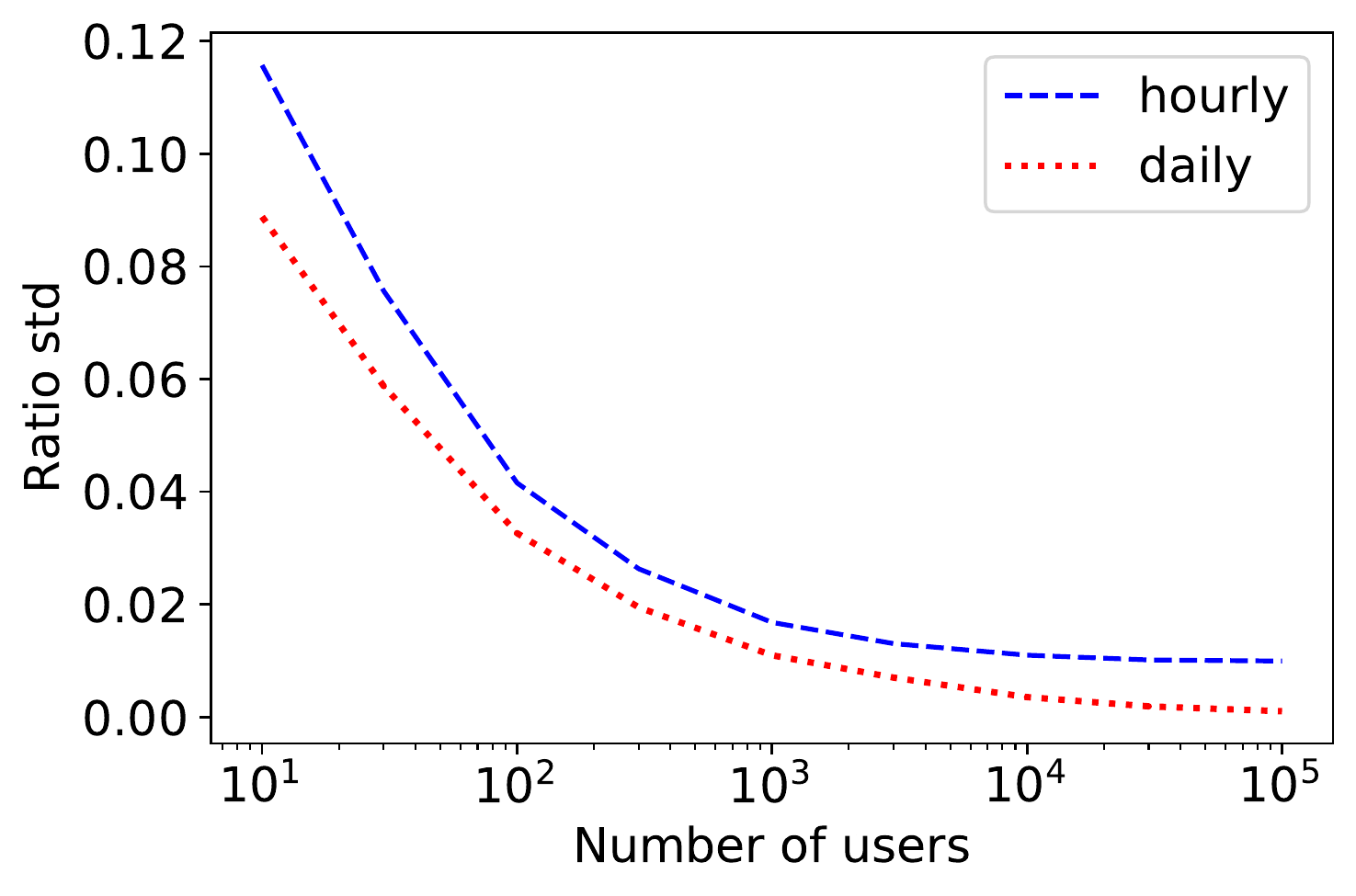}} 
\caption{The pattern of total bandwidth consumption for $100$ streaming users where bandwidth consumption is normalized to $\mathcal{C}/(24\times 30)$ (a) the same pattern for $10,000$ users (b), and the standard deviation of the ratio ``throttling/no throttling'' bandwidth consumption for different number of users (c). We have $\boldsymbol{V} = \{0, .1, ..., .9\}$, $\mathcal{C} = .8\sum_{i\in \mathcal{N}} R_ix_i$, $T= .3$, and accordingly, $r=.1$. }
\label{fig:streaming_diurnal}
\end{figure}

Figure \ref{fig:streaming_diurnal}(a,b) depict the bandwidth consumption of $100$ and $10000$ streaming users, respectively. The users initial $R_i$ is picked uniformly at random from the set $\boldsymbol{V} = \{.1, .2, ..., .9 \}$, and $\mathcal{C} = .8\sum_{i\in \mathcal{N}} R_i$. We set $T= .3$ and compute $r=0.1$ accordingly. Each user's start date is chosen uniformly at random from $30$ days, and the bandwidth consumption (y-axis) is normalized to the amount of supply (capacity) in each hour. We observe that when the users' start dates are distributed throughout the month, the bandwidth consumption with a non-limited capacity (no throttling) has a similar pattern compared to the case where the capacity is limited (throttling) where the daily average of bandwidth consumption in the former is reduced to accommodate the limited capacity.

Figure \ref{fig:streaming_diurnal}(c) summarizes the effect of throttling on the variability of aggregate bandwidth consumption across all users. The number of users the ISP supports is varied along the $x$-axis. For a given value of the x-axis, the corresponding y-axis value shows the variance of a curve comprised of the ratio of aggregate bandwidth consumption in the throttled system to that of the unthrottled system. A large y-axis value indicates that changes in user consumption over time differ greatly between throttled and non-throttled systems, whereas a low y-axis value indicates that throttling, while reducing overall consumption, does not affect the variability of this consumption much beyond the normal variability introduced by the combination of the underlying diurnal on-off nature that the users exhibit. We see that even for systems with small (tens) numbers of users, the difference in variance is less than $12\%$ and that this value drops quickly to 0 as the number of users is increased. In other words, simply the staggering of users' cycle start is effective at removing any additional variation in aggregate user consumption, and does not impose any additional ``feast'' and ``famine'' periods beyond what results from typical diurnal processes.

 \subsection{Streaming}
In this section, we focus on users who utilize the Internet for streaming. Users in these applications (e.g., when watching movies on Netflix), on average, utilize the Internet for a fixed fraction of the day. For simplicity, we assume that the user's streaming time only depends on the streaming content, not their available streaming rate. Hence, if an average user streams for a fraction $x_i$ of the billing cycle, they will stream for the same fraction if they get throttled, i.e., $y_i = x_i$. 

Additionally, since there is a finite number of streaming codecs in existence~\cite{grunenfelder1991characterization}, we let $\boldsymbol{V} = \{v_1,v_2, ..., v_k\}$ represent the rates offered across these codecs and assume that each user $i$ is able to stream a video at a set of rates $\boldsymbol{V}_i \subseteq \boldsymbol{V}$. It follows that a user's initial transmitting rate $R_i \in \boldsymbol{V}_i$ and that when throttled to a rate $r$, the user utilizes the largest value $v \in \boldsymbol{V}_i$ satisfying $v \le r$. 


Since $x_i = y_i$, and considering a simplified network consisting of streaming users, the bandwidth usage cannot surpass the network capacity. Therefore, we can rewrite Equation \ref{eq:uni_cap} for streaming users as:

 \begin{equation}
\mathcal{C} =
\sum_{i\in \mathcal{L}}R_ix_i + \sum_{i\in \mathcal{H}}(T+rx_i(1-\frac{T}{R_ix_i}))
\label{eq:C-streaming}
\end{equation}
in which the data allocation of a user $i\in \mathcal{L}$ is $B_i = R_ix_i$, and the data allocation of a user $i\in \mathcal{H}$ is $B_i =T+ r_ix_i(1-\frac{T}{R_ix_i})$.

\subsection{File download}



As mentioned previously, a file download user $i$ with the rate $R_i$, once throttled, may increase the fraction of time active from $x_i$ to $y_i \ge x_i$ in an attempt to still receive the same number of overall bits. Specifically, they would increase $y_i$ as much as possible to still obtain $R_i x_i$ bits during the billing cycle, i.e., such that $T + r y_i (1 - t_i) = R_i x_i$. Noting that $y_i \le 1$, we have:

\begin{equation}
 y_i = min\{\frac{R_ix_i-T}{r(\frac{R_ix_i-T}{R_ix_i})} = \frac{R_ix_i}{r}, 1\}  
\label{eq:y}\end{equation}

Equation \ref{eq:y} also tells us that $y_ir \leq R_ix_i$, which changes to an equality if $\frac{R_ix_i}{r} \leq 1$. This is intuitive since the users wish to have their average throttled consumption rate as close as possible to their average unthrottled consumption rate. 




We now describe a limitation on the choice of $y$ in file downloads.

{\lemma 
If $\mathcal{H} \ne \emptyset$, at least one user $i \in \mathcal{H}$ satisfies $y_i = 1$.\label{lemma:min-y}}

The proof can be found in the Appendix, and it can be used to show the following corollaries:
{\corollary
Any $i \in \mathcal{H}$ either has $y_i=1$ or $y_i = R_i x_i / r$.}

{\corollary \label{cor:only_small_y}
Any $i \in \mathcal{H}$ whose $y_i = R_i x_i / r$ consumes $R_i x_i$ bits per cycle.}
In other words, the only users affected by throttling are those users in $\mathcal{H}$ for whom $y_i = 1$.

Sine we assume file download users only care about the total bandwidth they receive, the ones for whom $r y_i = R_i x_i$ are unaffected by throttling, even in terms of regret as we see in Section \ref{sec:reg_download}. Therefore, for simplicity of presentation, we apply Corollary \ref{cor:only_small_y} and update the set $\mathcal{H}$ to only include the file download users whose $y_i=1$, i.e., the users who cannot achieve their unthrottled quantity of bits.

\section{Regret Model}\label{sec:reg_model}

Throttling increases the {\em regret} of a user, where this increase is amplified with a growing difference between the desired rate $R_i$ and the throttled rate $r_i$, as well as with a growth in the the amount of time the throttling is applied during the billing cycle. Since the ISP must choose a single threshold $T$ and throttled rate $r$ that applies to all users, the choice will affect the users' relative regrets. We assume that an ISP seeks to minimize the aggregate (sum of) regrets across all of its users.

\subsection{Fairness of Throttling and Regret}

Before continuing with the development of our regret model, we consider some issues of fairness with respect to throttling policy.  We make two observations when considering two users $i$ and $j$ where $i$ is ``more aggressive'' than $j$ in that $R_i x_i > R_j x_j$:

\begin{itemize}

\item The more aggressive user will hit the throttling threshold sooner in its cycle. Furthermore, the drop it experiences in rate will be more significant, leading to a larger loss of aggregate bit consumption within its cycle. Therefore, it is punished more severely along both \emph{time} and \emph{rate} dimensions.

\item While its punishment is more severe, its overall number of bits received will remain higher.

\end{itemize}

In this sense, this throttling mechanism is considered fair: the more aggressive you are, the more you lose, but your allocation of bits is never reduced below that of a less demanding user. If regret is measured as a function of this loss, then more aggressive users, while experiencing higher regret, will still receive more overall bits than their less aggressive counterparts.

There are many ways to define regret as an increasing function of both the fraction of time spent being throttled and the rate drop during throttling. In this paper, we consider a natural class of regret function for user $i$ as defined by:

\begin{equation}
\mathfrak{R}_i  =  (\frac{R_ix_i- ry_i}{R_ix_i})^\rho \times (1-\frac{T}{R_ix_i})^\tau \times 1_{\{i\in\mathcal{H}_j\}} =  (1-\frac{ry_i}{R_ix_i})^\rho \times (1-\frac{T}{R_ix_i})^\tau  \times 1_{\{i\in\mathcal{H}_j\}}  
\label{eq:ind_reg_old}
\end{equation}

Here, we multiply the regret due to the fraction of time spent throttled, by the regret due to the rate reduction relative to its unthrottled value. Note that the normalization factor $R_ix_i$ ensures that user regret is affected by their relative rate drop rather than the absolute value of it ($R_ix_i -ry_i$), which would be much more significant for high-demand users. Furthermore, the exponential parameters $\rho$ and $\tau$ can be varied to either increase or decrease the relative importance of the time factor and the rate factor. Hence, the aggregate regret for the ISP's consumers can be computed as:

\begin{equation}
\mathfrak{R}  =  \sum_{i\in \mathcal{H}}(1-\frac{ry_i}{R_ix_i})^\rho \times (1-\frac{T}{R_ix_i})^\tau   
\label{eq:regret_agg}
\end{equation}

\subsection{Regret minimization}
The main goal of ISP is to minimize the aggregate customers' regrets so that in a competitive non-monopolistic market, it would maintain the majority of them and do not lose them to its competitors. There are two criteria the ISP needs to account for to find the minimum regret in Equation \ref{eq:regret_agg}. First, some values of $r$ and $T$ may not be feasible to implement in the market. Second, the set of throttled users $\mathcal{H}$ is a function of $r$ and $T$ (\S \ref{sec:const-T-r}), where $\forall i\in \mathcal{H}: R_i \geq max(r, T)$. 

Given the capacity constraint of the system and the reverse relationship between $r$ and $T$, the throttling threshold can vary from $0$ to a maximum value, which we call $\hat{T}$. To compute $\hat{T}$, we first derive T as a function of $r$ using Equation \ref{eq:uni_cap}. We have:
\begin{equation}
T = \frac{\mathcal{C} - \sum_{i\in \mathcal{L}}R_ix_i - r\sum_{i\in \mathcal{H}} y_i}{\sum_{i\in \mathcal{H}}(1-ry_i/R_ix_i)}
\label{eq:T}
\end{equation}

As we increase $T$, $r$ decreases and vice versa. Therefore, if we plug $r = 0$ into Equation \ref{eq:T}, the throttling threshold gets its maximum value. Let's call this value $\hat{T}$, we have:

\begin{equation}
\hat{T} = \frac{\mathcal{C} - \sum_{i\in \mathcal{L}}R_ix_i}{\hat{H} }
\label{eq:T_max}
\end{equation}
 where $\hat{H}$ denotes the minimal cardinality of the throttled users. To find $\hat{H}$, we gradually shrink the set $\mathcal{H}$ (increase $T$), and compute $\hat{T}$ using the Equation \ref{eq:T_max} until $R_i > \hat{T}: \forall i\in \mathcal{H}$ (and $R_i \leq \hat{T}: \forall i\in \mathcal{L}$). We call this set $\hat{\mathcal{H}}$, and define $\hat{H} = |\hat{\mathcal{H}}|$. 

We need to make sure each throttled user $i \in \mathcal{H}$ has $R_i\geq r$, and the maximum ISP capacity can be consumed for minimizing the regret. Therefore, if for user $i \in \mathcal{H}$ has $R_i\leq r$, we move that user from the set $\mathcal{H}$ to $\mathcal{L}$ and recompute $r$ for the remaining users until no such user exists. This method can potentially lead to a circular approach and is computationally expensive. Therefore, in the following sections, we introduce a method to automatically compute the required sets and solve the regret minimization problem in video streaming and file downloads. 

\subsection{Regret analysis for streaming}\label{sec:stream}

Since for streaming users, we assume $y = x$,  
Equation \ref{eq:regret_agg} can be rewritten as: 
 \begin{equation} 
    \mathfrak{R} = \sum_{i\in \mathcal{H}} (1-\frac{r}{R_i})^{\rho} \times (1-\frac{T}{R_ix_i})^{\tau}
    \label{eq:reg-streaming}
\end{equation}

\begin{figure}[t]
\centering
\includegraphics[width=0.4\linewidth, angle=0]{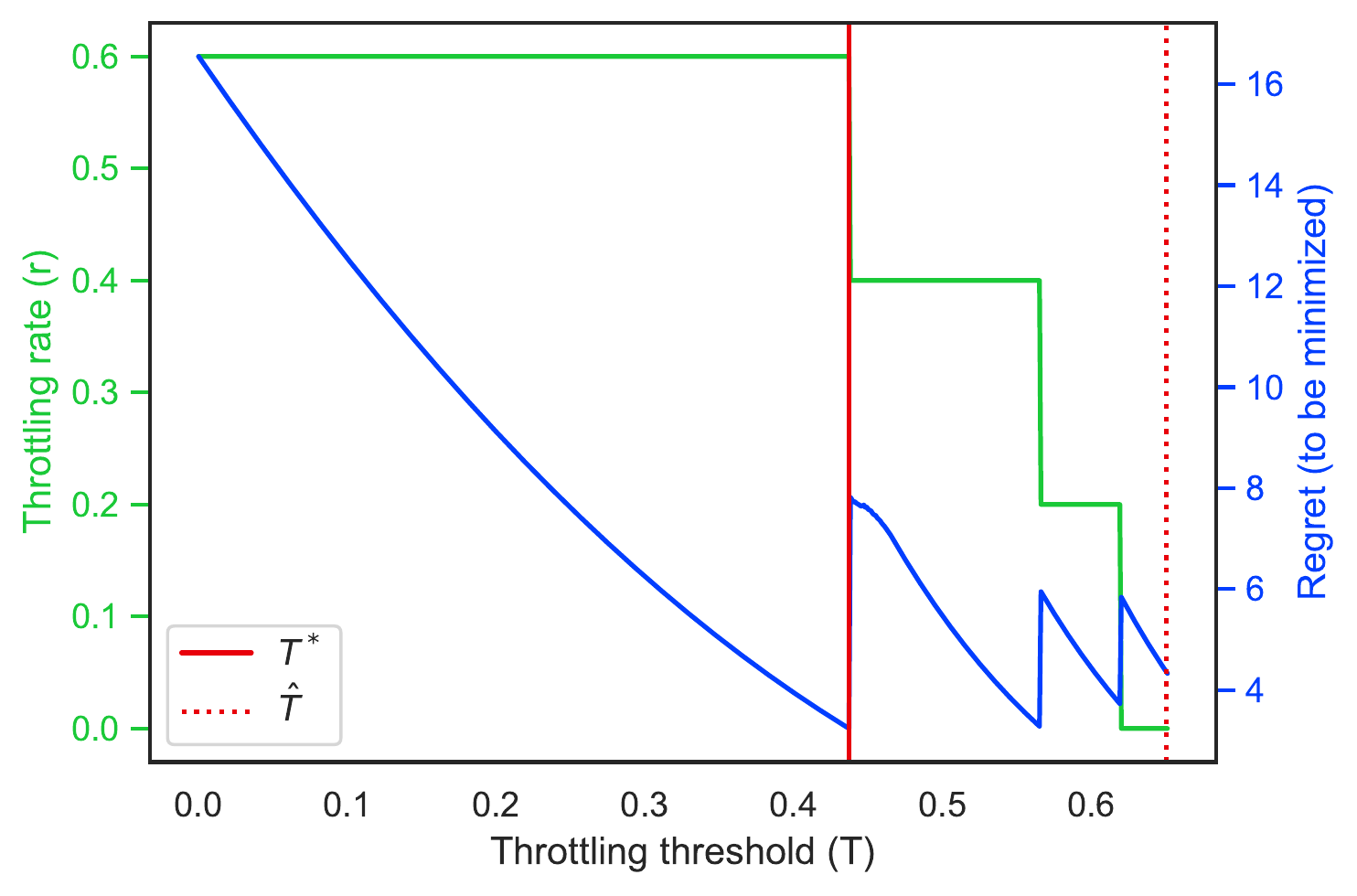}{} 
\caption{The total regret as a function of $T$, as well as its minimum. There are $5$ different codecs $\boldsymbol{V} = \{0.2,0.4,0.6,0.8,1.\}$, and $1000$ users with usage rates picked uniformly at random from $V$. We have $\mathcal{C} = 0.95 \sum_{i\in \mathcal{N}} R_ix_i$, and $\boldsymbol{X}$ comes from a uniform distribution.
}
\label{fig:reg_rT_video}
\end{figure}

Note that in Equation \ref{eq:reg-streaming}, $R_i$ and $r$ can only be chosen from the set of available video codecs, $\boldsymbol{V}$. To illustrate the regret function for streaming users, let's analyze an example with $5$ video codecs that users choose from, where all rates are normalized to $1$ and $\boldsymbol{V} = \{0.2,0.4,0.6,0.8,1.\}$. Also, let's assume there exist $1000$ users whose rates ($\boldsymbol{R}$) are chosen uniformly at random from $\boldsymbol{V}$, and $\boldsymbol{X}$ chosen uniformly at random from $\{0.01, 0.02, ...,1\}$. Figure \ref{fig:reg_rT_video} depicts $r$ as a function of $T$ (green curve), and total regret as a function of $T$ (blue curve). We assume the capacity is $95\%$ of bandwidth demand, i.e., $\mathcal{C} = 0.95 \sum_{i\in \mathcal{N}} R_ix_i$, and $\rho = \tau = 2$.

Let's start with the $r$ versus $T$ curve. As we increase the throttling threshold from $0$, given the capacity constraints of the ISP and Equation \ref{eq:C-streaming}, the highest achievable throttling rate that ISP can offer from $\boldsymbol{V}$ is $r = 0.6$. Note that at $T=0$, the chosen throttling rate $r = 0.6$ is not as high as the ISP cap can support, but is the highest achievable given the limited number of codecs. As $T$ grows to $\approx 0.42$, although $r$ stays the same, it becomes closer to the maximum rate the ISP can support. Simultaneously, we observe that when $T=0$, the regret curve starts from a maximum value since the entire ISP capacity is not utilized. As $T$ increases to $\approx 0.42$, the regret slowly drops since the chosen throttling rate $r$ becomes closer to the maximum that ISP can support.

Once $T$ becomes larger than $0.42$, the ISP can no longer support the rate of $r = 0.6$. Therefore, the throttling rate drops to the next available codec rate, i.e., $r = 0.4$. On the other hand, the total regret spikes up at $T \approx 0.42$ since again, the ISP cannot utilize its maximum capacity. Hence, with a slight increase of $T$ at this point, $r$ drops significantly, causing a spike in the regret value. As we keep increasing $T$, this behavior goes on, where $r$ versus $T$ forms a decreasing step function with discrete values of $r$ corresponding to different codec rates (green curve). On the other hand, $ \mathfrak{R} $ versus $T$ forms a Sharktooth function (blue curve). We continue until $T$ reaches its maximum value, i.e., $\hat{T}$. The minimum regret is shown by the vertical line of $T= T^*$ which occurs in a local minimum of the Sharktooth regret function.

In general, regardless of the parameters $\rho$ and $\tau$, the minimum regret may occur at any valley of the Sharktooth regret function (i.e., any jump point of $r$ v.s. $T$ step function). Intuitively, the reason is that only at those points, the maximum available ISP capacity is allocated to the users. Therefore, we only need to investigate $|\boldsymbol{V}|$ many points (where $r\in \boldsymbol{V}$) to find the global minimizer of regret. We use algorithm \ref{alg:reg_min_stream} to compute $T^*$ and minimum regret $\mathfrak{R}^*$ for streaming users.

\begin{algorithm}[t]
\DontPrintSemicolon
\caption{Regret minimization for streaming}
\label{alg:reg_min_stream}
\SetKwProg{generate}{Function \emph{generate}}{}{end}
Define the regret function as Equation \ref{eq:regret_agg}.\\
Compute $\hat{T}$ using Eqn. \ref{eq:T_max}.\\
$y_i \leftarrow x_i \forall i\in \mathcal{N}$.\\
Initialize $T^* \leftarrow 0$  and $\mathfrak{R}^*\leftarrow \infty$. \\
\ForAll{ $r \in \boldsymbol{V}$ }{
    Compute $T$ using $T \leftarrow binarySearch(0,\hat{T} )$.\\
    Compute the resulting $\mathfrak{R}$ using Equation \ref{eq:reg-streaming}. \\ 
    \If{  $\mathfrak{R}$ < $\mathfrak{R}^*$}{$T^* \leftarrow T$  and $\mathfrak{R}^*\leftarrow \mathfrak{R}$.\\}
}
return $T^*$ and $\mathfrak{R}^*$\\
  \SetKwFunction{Fbin}{binarySearch}
  \SetKwProg{Fn}{Function}{:}{}
  \Fn{\Fbin{$lo$, $hi$}}{
  Initialize $\epsilon$ to some small number.\\
   \While{$lo<hi$}{
    $mid \leftarrow (lo+hi)/2.$\\
    $\mathcal{H} \leftarrow \{i$ for $i\in \mathcal{N}$ if $R_i > max(r, mid) \}$. \\
    $\mathcal{L} \leftarrow \mathcal{N} - \mathcal{H} $.\\
    Compute $T$ using Equation \ref{eq:T}. \\
    \If{ $|mid-T| < \epsilon$ }{return $T$.\\}
    \ElseIf{ $mid < T$}{$lo \leftarrow mid$.\\}
    \ElseIf{$mid > T$}{ $hi \leftarrow mid$.\\}
    }
  }
\end{algorithm}


\subsection{Regret analysis for file downloads}\label{sec:reg_download}

In this section, we introduce the notion of kickin and kickout points to facilitate the regret analysis for file downloaders. If $T<R_i$, user $i$ may get throttled until $T$ grows to be larger than $R_i$. Let's define kickout points as any value of $T_0$ for which if $T$ changes from $T_0 - \epsilon$ to $T_0 + \epsilon$ for $\epsilon \to  0$, at least one user $i$ stops getting throttled. Hence, the kickout points could be defined as $T_0 =\{ R_i: \forall i\in \mathcal{N} \text{ if } R_i<\hat{T}$\}, and for each point in that set, a user is removed from the set $\mathcal{H}$ and added to the set $\mathcal{L}_H$. Furthermore, we define kickin point to be any $T_0$ for which if $T$ changes from $T_0 - \epsilon$ to $T_0 + \epsilon$ for $\epsilon\to 0$, at least one user $i$ starts getting throttled (is added to the set $\mathcal{H}$). One question to ask is that does user $i$ get throttled for any $T<R_i$? The answer is no. If $T$ is too small, due to the reverse relation of $r$ and $T$, $r$ would be large. When $r \geq R_i$, user $i$ cannot get throttled to a higher rate. As $T$ grows, $r$ decreases until at some point $r < R_i$, and if we still have $T<R_i$, user $i$ gets throttled. In other words, a user is removed from the set $\mathcal{L}_R$ and is added to the set $\mathcal{H}$. The kickin points can occur for any $T_0$ where the respective throttling rates come from the set $r_0 =\{ R_i: \forall i\in \mathcal{N}\}$. We plug the set $r_0$ into Equation \ref{eq:T}, and if the result meets $T< min(\hat{T}, R_i)$, we add the computed $T$ to the set of kickin points. 

\begin{figure}[t]
\centering
\hbox{\hspace{26ex}\includegraphics[width=.65\linewidth, angle=0]{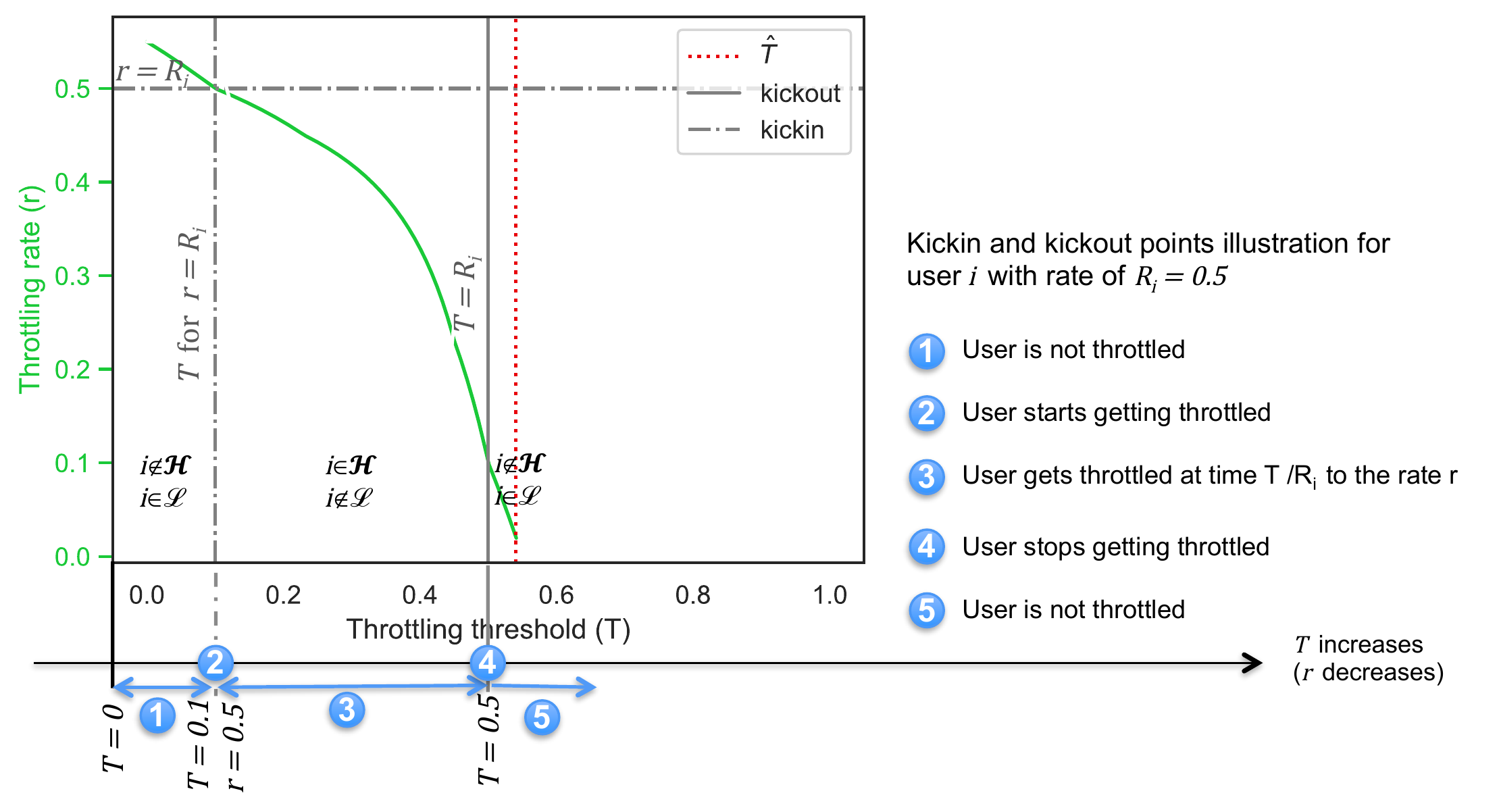}{} } 
\caption{The kickin and kickout point of a particular user $i$ with $R_i = 0.5$. The $r$ versus $T$ graphs represents a distribution with $\boldsymbol{R} = [0.3, 0.45 , 0.5, 1.]$ and $\mathcal{C} = 1.8$.
}
\label{fig:kickin_kickout}
\end{figure}

Figure \ref{fig:kickin_kickout} illustrates an example of the $r$ versus $T$ curve and shows when user $i$ ($R_i = .5$) gets kicked in and out of the set $\mathcal{H}$ as we increase $T$. When $T<0.1$ (1), we have $R_i < r$ and user $i$ not throttled; since she may not get throttled to a higher rate. When $T=0.1$ (2), we have $R_i = r$ and the user starts getting throttled (gets added to the set $\mathcal{H}$). When $0.1<T<0.5$ (3), we have $R_i>max(r, T)$ and the user is in the set $\mathcal{H}$. The user in this case experiences throttling after using the limit of $T$ at time $T/R_i$ of her cycle, and her new rate is reduced to $r$. Finally, when $T=0.5$ (4), we have $T=R_i$, and the user is moved from the set $\mathcal{H}$ to $\mathcal{L}$ and is no longer throttled for any $T>R_i$ (5).

Given $N$ many users, there exist a maximum of $2N$ many kickin/kickout points. This number could potentially be smaller than $2N$ since each computed kickin/kickout point needs to be smaller than $\hat{T}$. Between any two consecutive $\{kickins \cup kickouts\}$ points, the set $\mathcal{H}$ remains intact and the regret function is convex, making it easier to compute the minimum aggregate regret on that interval. Our regret model for all intervals then introduces a semi-convex regret function with a global minimum, where our goal is to find the optimization point. We define a semi-convex function as below:
 
 {\definition[semi-convex regret function] The term semi-convex describes a regret function that is convex for each interval of $T$ where the set of throttled users $\mathcal{H}$ is unchanged.
\label{def:semi-convex}}\\

After we compute $\mathcal{H}$ and $\mathcal{L}$ through the sets of kickin and kickout points, for any interval where $\mathcal{H}$ is constant, we can compute the local minimum. In Equation \ref{eq:regret_agg}, $\rho = \tau$ implies that users give the same weight to the fraction of rate drop versus the fraction of time they are throttled for, which is not far from reality. Therefore, although our general model can consider any values of $\rho$ and $\tau$, from now on, we assume $\rho = \tau$ and show this assumption adds the following property to our regret function (proof in the appendix). 

{\theorem For file download users, if $\rho=\tau\geq 2$ in Equation \ref{eq:regret_agg}, the minimizer of regret given a constant $\mathcal{H}$ is independent of $\rho$ for all $\rho\geq 2$ and is computed as:
\begin{equation}
    T^* = \frac{H - \sqrt{H^2 - (\mathcal{C} - \sum_{i\in \mathcal{L}}R_i)\times (\sum_{i\in \mathcal{H}}\frac{1}{R_i})} }{\sum_{i\in \mathcal{H}}\frac{1}{R_i}}
\label{eq:T_min}
\end{equation}
\label{lemma:min_regret}}

\begin{figure}[t]
\centering
\includegraphics[width=0.4\linewidth, angle=0]{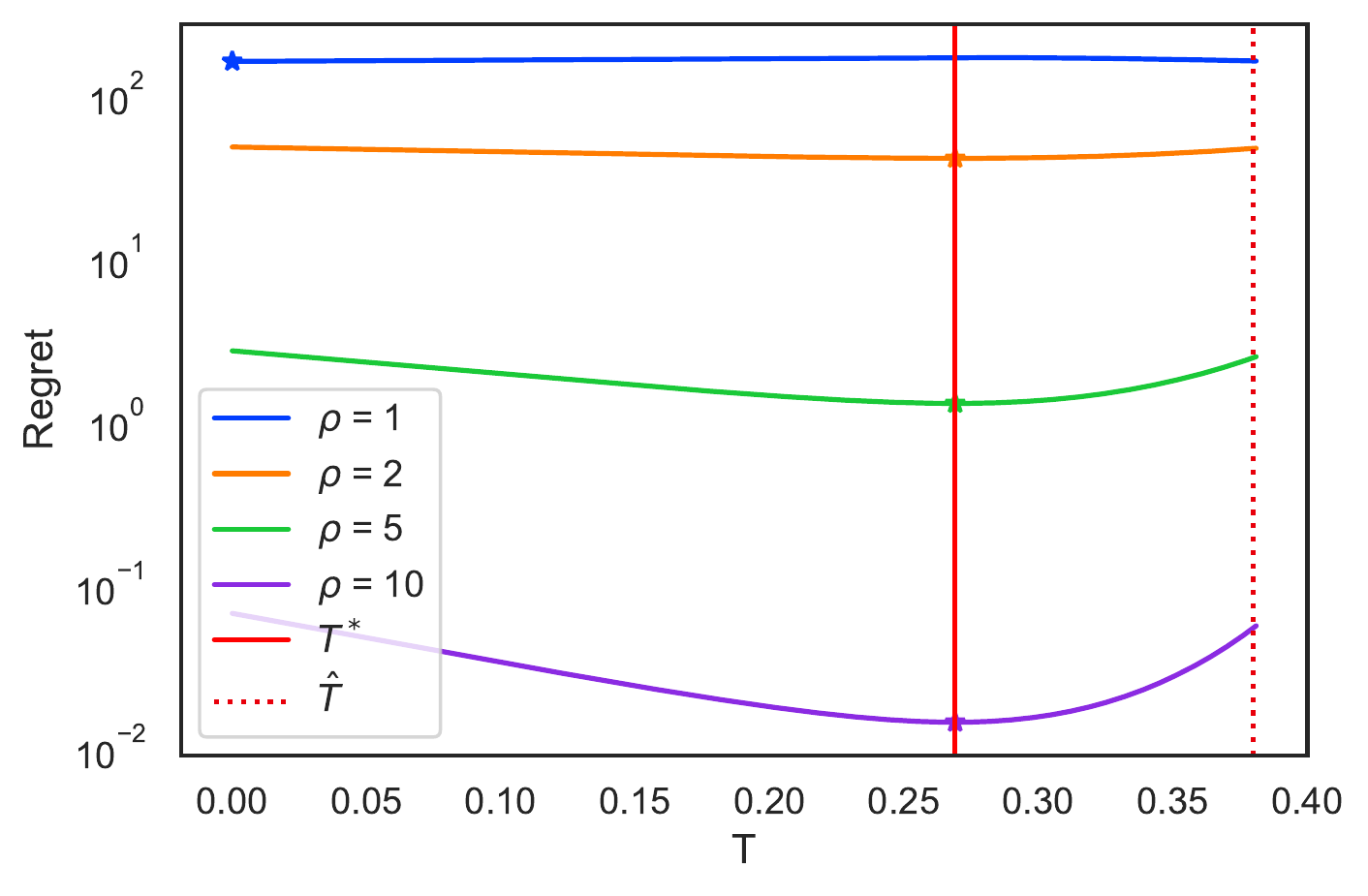}{} 
\caption{The total regret as a function of $T$, as well as its minimum for different $\rho$, where $\tau = \rho$. There are $1000$ users with log-normal distribution of $\boldsymbol{R}$ with $\sigma = 0.25$ and $\mu = 1$, and $\mathcal{C} = 0.8 \sum_{i\in \mathcal{N}} R_ix_i$.
}
\label{fig:reg_vs_T}
\end{figure}

 Figure \ref{fig:reg_vs_T} depicts the regret (log-scale) versus throttling threshold $T$ for different values of $\rho$, as well the minimization point of the regret function, $T^*$. We observe that for all $\rho =\tau \geq 2$, the regret minimizer $T^*$ does not change. 
Note that if $ 1 < \rho < 2$, the regret function is not necessarily semi-concave or semi-convex (As seen in the Figure and also included in the proof of Theorem \ref{lemma:min_regret}). Therefore, our focus is on $\rho=\tau\geq 2$.
 

A step-by-step procedure of computing the kickin/kickout points, and then finding $T^*$ is presented in Algorithm \ref{alg:reg_min}. We use the regret function in Equation \ref{eq:regret_agg} for file downloads, assuming $\rho=\tau \geq 2$.

\begin{algorithm}[t]
\DontPrintSemicolon
\caption{Regret minimization for file downloads}
\label{alg:reg_min}
\SetKwProg{generate}{Function \emph{generate}}{}{end}
Define the regret function as Equation \ref{eq:regret_agg}.\\
Compute $\hat{T}$ using Eqn. \ref{eq:T_max}.\\
Initialize an empty sets of $kickins$ and $kickouts$. \\
Initialize a hashmap of $kickins$ to corresponding $R_i$ as $kickins2R_i$. \\
\ForAll{ $i \in \mathcal{N}$ }{
    $r\leftarrow R_i$.\\
    Compute $T$ using Eqn. \ref{eq:T}.\\
    \If{ $T$ < $\hat{T}$}{Insert $T$ into the set of $kickins$.\\
    Insert $R_i$ into the hashmap of $kickins2R_i[T]$\\}
    \If{$R_i < \hat{T}$}{Insert $R_i$ into the set $kickouts$.\\}
 
}
Sort the set $\{kickins \cup kickouts\}$. \\
Define global minimum of Regret to be $\mathfrak{R^*} \leftarrow \infty$.\\
Define $T^*$ to minimize of the Regret.\\
 \ForAll{consecutive $a,b$ in the set $\{kickins \cup kickouts\}$}
  {
        Define interval $[a,b)$.\\ 
        \If{$a\in kickins$}{ Insert the corresponding $R_i$ from the hashmap $kickins2R_i[a]$ into $\mathcal{H}$. \\}
        \ElseIf{$a\in kickouts$}{Remove $a$ from  $\mathcal{H}$.\\}
        Compute $T$ minimizing $\mathfrak{R}$ in $[a,b)$ using Eqn. \ref{eq:T_min}.\\
        \If{$\mathfrak{R} < \mathfrak{R^*}$}{ $\mathfrak{R^*} \leftarrow \mathfrak{R}$.\\
        $T^* \leftarrow T$.\\}
    }
return $T^*$ and $\mathfrak{R^*}$
\end{algorithm}

\begin{figure}[t]
\centering
\subfigure[Total regret]{\includegraphics[width=0.325\linewidth, angle=0]{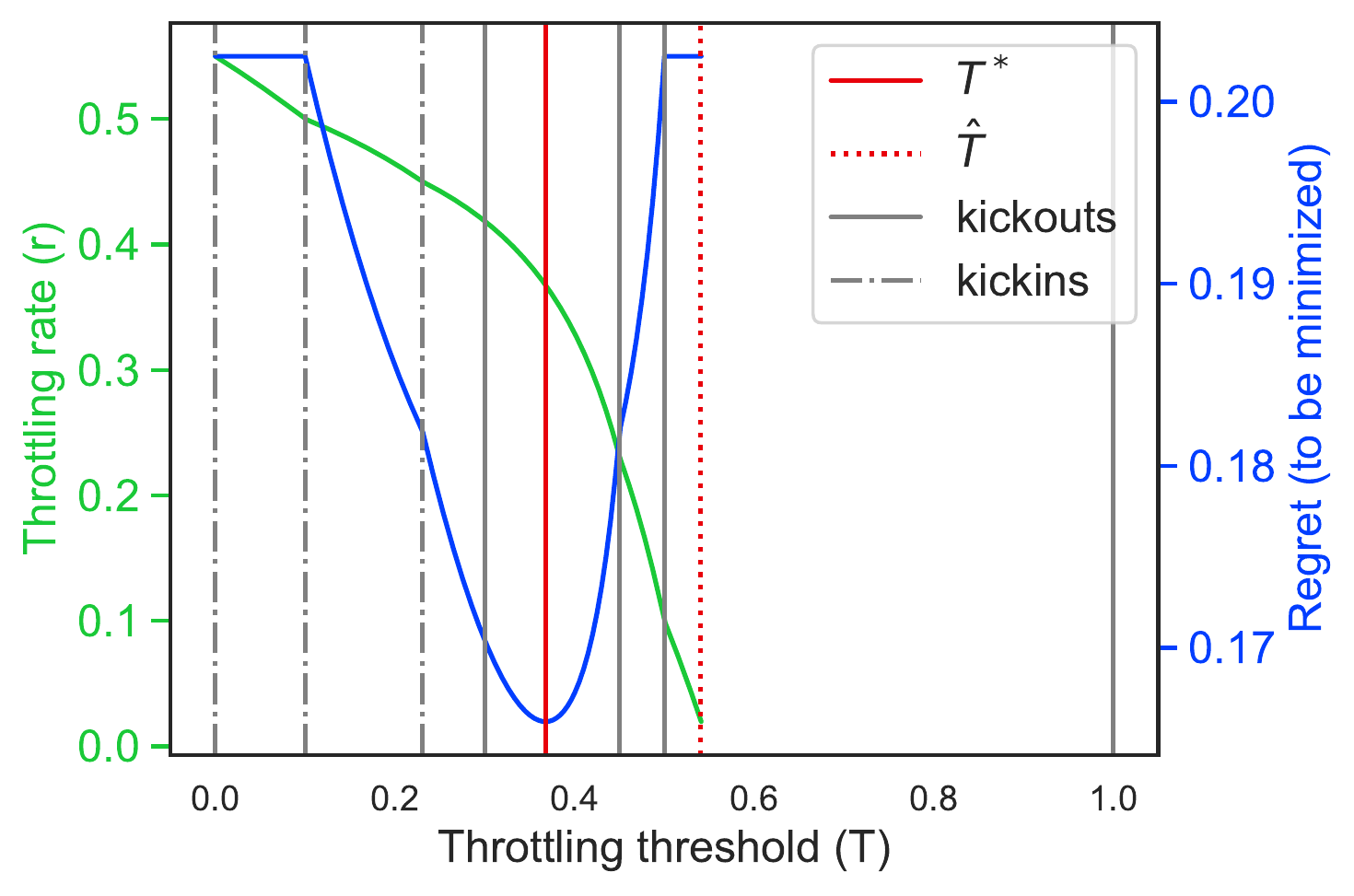}} 
\subfigure[Users' individual regrets]{\includegraphics[width=0.325\linewidth, angle=0]{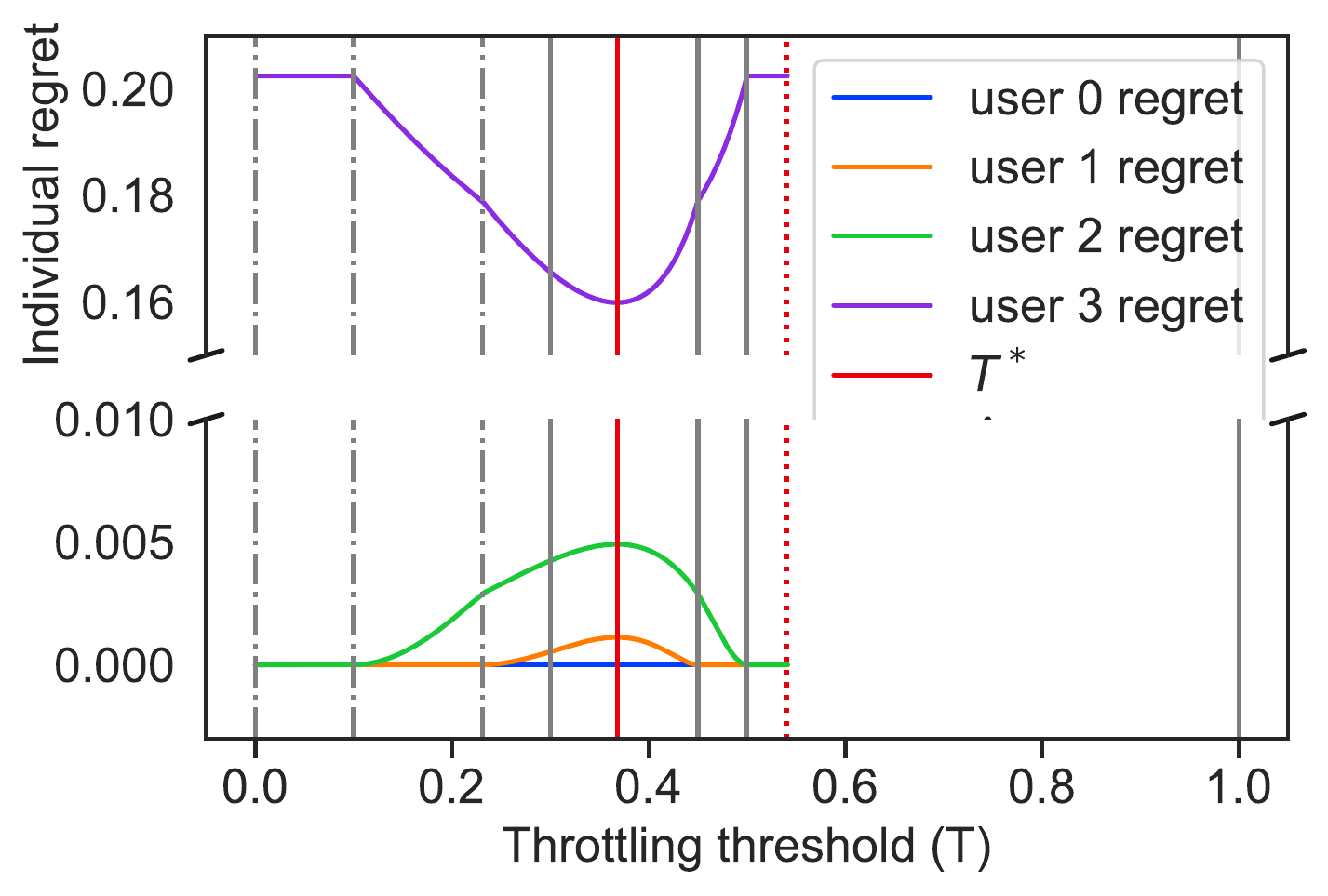}} 
\subfigure[Users' individual bandwidth allocations]{\includegraphics[width=0.325\linewidth, angle=0]{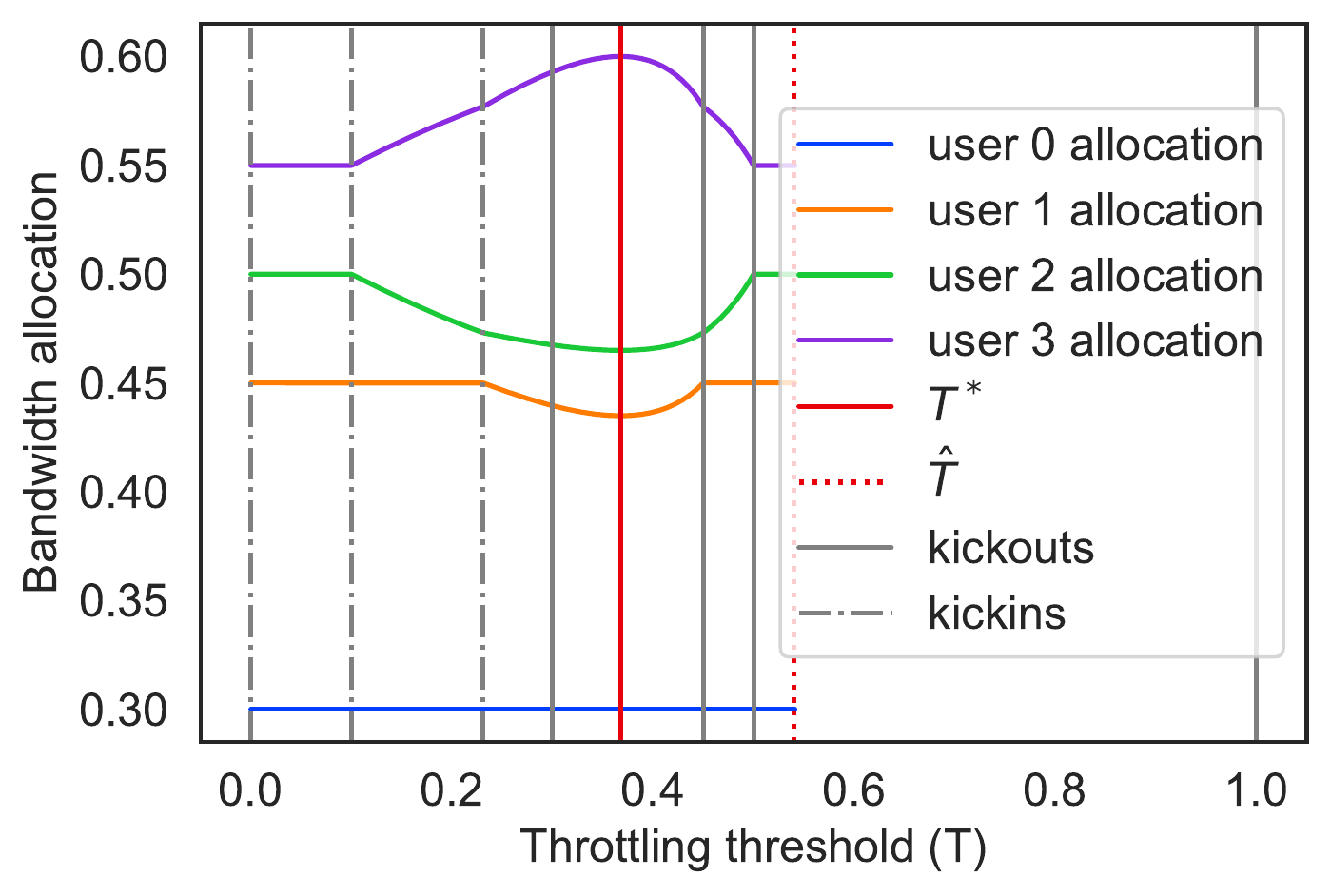}} 
\caption{Total regrets, individual regrets, and individual bandwidth allocations. We have $[R_1, R_2, R_3, R_4] = [0.3, 0.45 , 0.5, 1.]$, $\mathcal{C} = 1.8$, and $\rho = \tau = 2$}
\label{fig:3users}
\end{figure}

To elaborate how this algorithm works, Figure \ref{fig:3users} depicts the regrets and allocations for four users with $\boldsymbol{R}$ selected from $(0,1]$, $\mathcal{C} = 0.8\sum_{i\in \mathcal{N}}R_i$, and $ 0\leq T\leq \hat{T} $. In all three sub-graphs, $T^*$, kickin, and kickout points are shown by a vertical red line, solid gray lines, and dashed gray lines, respectively. We define each interval to be formed by two consecutive values of $T$ in the sorted set of kickin and kickout points, i.e, $(T_i, T_{i+1}] : T_i, T_{i+1} \in sorted\{kickins\cup kickouts\}$, where at each interval $\mathcal{H}$ is unchanged. Figure \ref{fig:3users}(a) depicts the aggregate regret, as well as $r$ versus $T$ curve. In Figure \ref{fig:3users}(b), we observe that user $3$ with the highest rate always has the maximum regret, which is minimized at $T^*$. Furthermore, the difference between its regret and the other regrets is minimum at $T^*$. Also, user $0$ has no regret since she does not get throttled at all due to its low $R_0$. Figure \ref{fig:3users}(c) shows that user $3$ with the highest $R_i$, has the highest bandwidth allocation among users. However, it is \emph{always} lower than its desired bandwidth, $R_3$. On the other hand, users $1$ and $2$ have their bandwidth allocations lower than $R_1$ and $R_2$ only for limited intervals of $T$, respectively. We introduce the following lemma to theoretically verify this finding, where its proof is available in the appendix.

{\lemma For any two throttled users $i, j \in \mathcal{H}$ where $R_i > R_j$, we have $B_i > B_j$ and $\mathfrak{R}_i > \mathfrak{R}_j$. 
\label{lemma:ind_reg}}

Note that although Lemma \ref{lemma:ind_reg} indicates a higher bandwidth allocation for the high-demand users compared to low-demand ones, their regrets are still higher than that of low-demand users. This is an important ``fairness'' property, i.e., a higher-rate user is not ``punished'' unfairly, in terms of both regret and allocation, for its (presumably legitimate) behavior. Yet, they do not get a significantly higher allocation than other users and their demand is capped. 



Based on what we discussed for \emph{file downloaders}, to compute the regret function, we have $\rho = \tau$ and only consider user $i$ in $\mathcal{H}$ if $y_i = 1$. Therefore, the regret function in Equation \ref{eq:regret_agg} will adopt a symmetric form with respect to both $r$ and $T$. The following lemma proves that this symmetric look will result in $r^* = T^*$ at the optimal point, which is then used in the proof of the main theorem of this section, i.e., Theorem \ref{theorem:interval_reg}.

{\lemma At the optimal point $T^*$, we have $r = T^*$.\label{lemma:r-equals-T}}

Previously we showed through Theorem \ref{lemma:min_regret} that when $\tau = \rho\geq 2$, the local regret minizer $T$ \emph{in each interval} of $\{kickins \cup kickouts\}$ (where $\mathcal{H}$ is unchanged) is independent of $\rho$. The next step is to prove that the interval that contains $T^*$ is also independent of $\rho$. Therefore, the global regret minimizer, $T^*$ is also independent of $\rho$ for any $\tau = \rho\geq 2$ (proof in the appendix). 

{\theorem The interval where the regret (Equation \ref{eq:regret_agg}) is minimized is independent of the regret function exponent $\rho$ when $\rho=\tau\geq 2$. \label{theorem:interval_reg}}

\section{multiple tiers}\label{sec:multiple_tiers}
In this section, we extend our analysis to multiple {\em tiers}, where users who use the same ISP plan form the same tier. Each tier $j$ is assigned a threshold $T_j$ and post-throttling rate $r_j$, and a user's tier choice depends on ($T_j, r_j$, $p_j$), where $p_j$ is the per-cycle price associated with tier $j$. We assume each tier $j$ is allocated the capacity $\mathcal{C}_j$ such that $\sum_j \mathcal{C}_j = \mathcal{C}$. We then investigate ISP's decision on each tier's capacity, and users' decisions on selecting the tier to minimize their individual regrets. While this section focuses on file download, we can easily extend it to the streaming data type. As future work, we can build a hybrid of streaming and file download, where each data type can form a separate tier, and we may introduce dummy users to capture the ones who utilize both data types.

We first analyze two tiers in Section \ref{sec:two_tiers} and show that there exist multiple Nash equilibria for different allocations of the ISP's capacity to the tiers. Among those equilibria, the ISP can choose an allocation that makes the average total regret minimized. We then show in Section \ref{sec:three-tiers} how we can extend this to more than two tiers, where we use a Non-Linear Program (NLP) to find the optimum bandwidth allocation so that the ISP can minimize the average total regret. 

\subsection{Two tiers}\label{sec:two_tiers}
Suppose the ISP provides multiple tiers for the users, each with its unique $T_j$ and $r_j$. Let's start with two tiers. The users who pay the price $p_1$ are throttled to rate $r_1$ after the bandwidth usage of $T_1$, and the users who pay the price $p_2$ are throttled to rate $r_2$ after the bandwidth usage of $T_2$. Note that we assume each tier has an exogeneous price determined by the market, and for simplicity, the ISP has no control over it. Without loss of generality, let's assume $p_1 < p_2$, where $\mathcal{L}_1$ and $\mathcal{L}_2$ are the set of unthrottled users in tiers $1$ and $2$, and $\mathcal{H}_1$ and $\mathcal{H}_2$ are the set of throttled users in tiers $1$ and $2$, respectively. Furthermore, let $\kappa$ denote a positive constant to model the impact of price on each user's decision, i.e., the higher $\kappa$ is, the more conservative users are regarding paying bandwidth prices. The total bandwidth consumption must satisfy the following equation:

\begin{equation}
\mathcal{C}  =  \sum_{j\in \{1,2\}}\left( \sum_{i\in \mathcal{L}_j}R_ix_i +  \sum_{i\in \mathcal{H}_j}T_j + r_j (1-T_j/R_ix_i)\right)
\end{equation}
To account for the tiers' prices in users' decisions, user $i$'s regret choosing tier $j$ is computed as:
\begin{equation}\label{eq:2tier_ind_reg}
\mathfrak{R}_i =\kappa  p_j + 
    (1-\frac{r_1}{R_ix_i})^{\rho} \times (1-\frac{T_1}{R_ix_i})^{\tau}\times 1_{\{i\in\mathcal{H}_j\}}
\end{equation}
where the first term models the tier's price in user's regret and hence their decision, and the second term is similar to the regret model we had in Section \ref{sec:reg_model}. Note that in a $1$-tier model in the previous section, if we add the price term as in $\sum_{\mathcal{N}} p \times \kappa $, it would be constant with no impact on the regret minimization and hence can be removed. Therefore, given Equation \ref{eq:2tier_ind_reg}, the total regret function for a two-tiered ISP is calculated from the following equation.

\begin{equation}\label{eq:2tier_reg}
    \mathfrak{R} = \sum_{j\in \{1,2\}}\left(\sum_{i\in \mathcal{H}_j} (1-\frac{r_j}{R_ix_i})^{\rho} \times (1-\frac{T_j}{R_ix_i})^{\tau}+ \sum_{\mathcal{L}_j\cup \mathcal{H}_j}  \kappa p_j \right)
\end{equation}

To find a regret minimization model for two tiers, we assume $\mathcal{C}_1$ and $\mathcal{C}_2$ are the bandwidth allocations of tiers $1$ and $2$, respectively, and $\mathcal{C}_1+\mathcal{C}_2 = \mathcal{C}$. After the ISP allocates bandwidth to tier $j$, we can compute the $T_j$ minimizing the regret given tier $j$'s users. Specifically, for streaming, we use Algorithm \ref{alg:reg_min_stream}, and for file download, we use Algorithm \ref{alg:reg_min}. Having the minimum regret of each tier, we can then derive the \emph{total minimum regret} as a summation of them. The ISP, hence, could iteratively change the allocation between $\mathcal{C}_1$ and $\mathcal{C}_2$ and compute the minimum total regret until this summation is minimized. Each user may also decide whether to switch tiers based on $T_j$, $r_j$, and $p_j$ for each tier. Assuming users are rational, each user's goal is to minimize their individual regret. Therefore, we build a game-theoretic model to consider the decisions of the ISP as the leader, and the users as followers, which resembles a Stackelberg game \cite{von2010market}.

\begin{figure}[t]
\centering
\includegraphics[width=0.4\linewidth, angle=0]{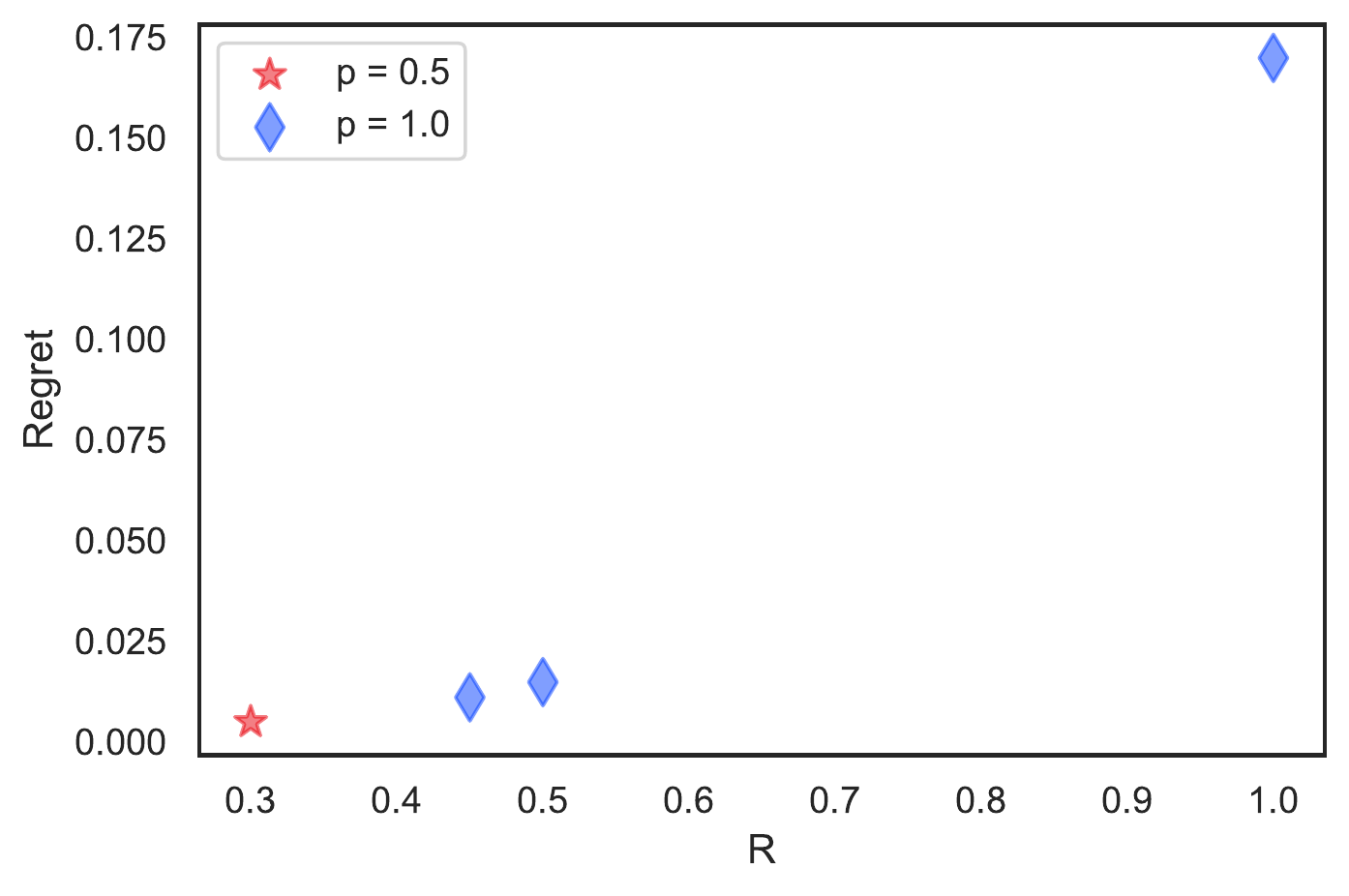}{} 
\caption{A Nash equilibrium for $\boldsymbol{R} = [0.3, 0.45, 0.5, 1.]$, $(p_1, p_2)$ = (0.5, 1.), $\mathcal{C} = 1.8$, and $\kappa = 0.01$. After the Equilibrium, $[\mathcal{C}_1, \mathcal{C}_2] =[0.3, 1.5]$ and $[T_1, T_2 ]= [r_1, r_2 ]= [0.3, 0.37] $. We observe that the user with lowest values of $R_i$ choose tier $1$, and the other users choose tier $2$.
}
\label{fig:2tier_eq}
\end{figure}

To illustrate how this model works, we use an example of $4$ users with normalized rates of $\boldsymbol{R} = [0.3, 0.45, 0.5, 1.]$\footnote{We choose this arbitrary example to capture a small market with the majority consuming a rate in the middle, and the minority have very low or very high rates, but our analysis works for any distribution of $\boldsymbol{R}$.}. We also assume the network capacity is $80\%$ of the total data usage, i.e., $\mathcal{C} = 1.8$, and the ISP has two tiers with exogenous prices normalized to $1$, where $p_1 = 0.5$ and $p_2 = 1$. One of the resulting Nash equilibria is shown in Figure \ref{fig:2tier_eq}, where the star markers indicate the users who choose tier $1$ and the diamond markers indicate the users who choose tier $2$.

\begin{figure}[t]
\centering
\includegraphics[width=.7\linewidth, angle=0]{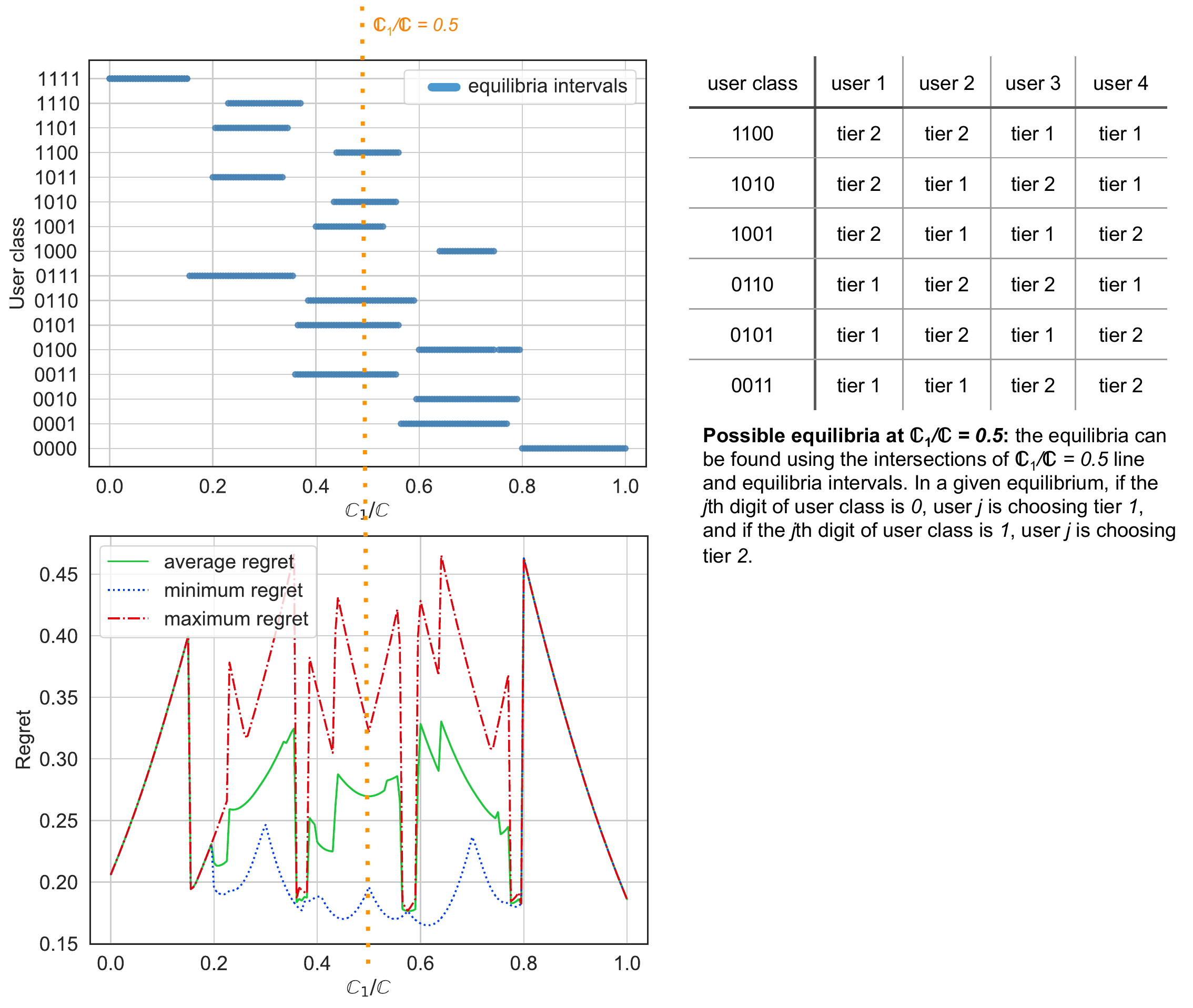}{} 
\caption{All possible Nash equilibria for $\boldsymbol{R} = [0.3, 0.45, 0.5, 1.]$, $(p_1, p_2)$ = (0.5, 1.), $\mathcal{C} = 0.8\sum_{i\in \mathcal{N}} R_i = 1.8$, and $\kappa = 0.01$. The x-axis shows the fraction of capacity assigned to tier $1$, and the y-axis in the top and bottom graph show the class of users and total regret, respectively. The bottom figure shows the impact of tier capacity assignment on average regret, minimum regret, and maximum regret across all equilibria for any given $\mathcal{C}_1/\mathcal{C}$.
}
\label{fig:eq_intervals}
\end{figure}

Figure \ref{fig:2tier_eq} shows only one possible equilibrium under the given configuration. In the case of multiple equilibria, the ISP has multiple options of allocating its capacity to different tiers. To further demonstrate this, we present every equilibrium of the same distribution in Figure \ref{fig:eq_intervals}. Given $4$ different users, there are $2^4 = 16$ ways they can select from two tiers, i.e., there can exist $16$ different classes of users who can behave differently. We assign a $4$-digit binary ID to each class ranging from $0000$ to $1111$. If digit $i$ of this ID is $0$, we assume user $i$ has picked tier $1$, and if digit $i$ is $1$, user $i$ has picked tier $2$. These different classes are depicted in the $y$-axis of Figure \ref{fig:eq_intervals} (top). The $x$-axis represents $\boldsymbol{C}_1/\boldsymbol{C}$ which is being varied from $0$ to $1$, and $\boldsymbol{C}_2 = \boldsymbol{C}-\boldsymbol{C}_1$. The markers on the graph depict the Nash equilibria for each $\boldsymbol{C}_1/\boldsymbol{C}$, i.e., they show what different classes of users result in the equilibria. For instance, suppose the ISP allocates $50\%$ of its capacity to each tier ($\boldsymbol{C}_1/\boldsymbol{C} = 0.5$). This allocation is shown by a vertical orange line in Figure \ref{fig:eq_intervals}. There are $6$ classes that have an intersection with the orange line, which can result in equilibria and are shown on the table of Figure \ref{fig:eq_intervals}.

Since for each $\boldsymbol{C}_1/\boldsymbol{C}$ there may exist multiple Nash equilibria, the ISP can decide on the capacity allocation for each tier based on its own goals. For instance, the ISP's goal could be minimizing the maximum regret among different equilibria, minimizing the minimum regret, minimizing the average regret, or a combination of all. Figure \ref{fig:eq_intervals}(bottom) demonstrates how the ISP could decide on assigning its capacity to different tiers. As before, the x-axis represents $\mathcal{C}_1/\mathcal{C}$ ($\mathcal{C}_2 = \mathcal{C}/\mathcal{C}_1$). In this example, we observe that around $\boldsymbol{C}_1/\boldsymbol{C} = 0.38$, $\boldsymbol{C}_1/\boldsymbol{C} = 0.58$, and $\boldsymbol{C}_1/\boldsymbol{C} = 0.79$ the average regret, minimum regret, and maximum regret all have small values. Therefore, each of these points could be a good split point that the ISP could choose, resulting in the corresponding possible user classes depicted on the top graph.


\subsection{Three tiers and more}\label{sec:three-tiers}
When the number of tiers increases, while it is possible to use a similar technique as in Section \ref{sec:multiple_tiers}, the computations become exponentially more expensive and finding the regret minimizer will become more complicated. Therefore, to find the optimal allocation of the ISP's capacity among different tiers, we propose solving a non-linear program (NLP). We first assume users are fixed (they do not change their initial tiers), then the ISP could find the optimal allocation among tiers. Users can then change their tiers to minimize their individual regrets. After that, the ISP can adjust its allocation to minimize the aggregate regret. The last two steps are repeated until the system reaches equilibrium, or until a counter we define maxes out, in which case we cannot detect any equilibria.

To solve the NLP, we use \emph{optimize.minimize} from the Scipy package in python \cite{NLP}, with the solver \emph{SLSQP}. This solver uses Sequential Least-Squares Programming to minimize a function of several variables with any combination of bounds, equality, and inequality constraints \cite{NLP, kraft1988software}.

As an example, we depict an ISP with three tiers and the normalized tier prices of $(p_1, p_2, p_3) = (0.5, 0.75, 1.)$, and assume the case of file downloads. This method can be extended to any number of tiers, as well as to streaming. We first set the bounds on each tier to $[R_{min}, R_{max}]$, where $R_{min}$ is the minimum bandwidth rate among users, and $R_{max}$ is the maximum of that. The goal of our optimizer is to minimize the regret function below:

\begin{equation}\label{eq:3tier_reg}
    \mathfrak{R} =  \sum_{j\in \{1,2,3\}}\left( \sum_{i\in \mathcal{H}_j}  (1-\frac{r_j}{R_ix_i})^{\rho} \times (1-\frac{T_j}{R_ix_i})^{\rho}+ \sum_{i\in \mathcal{L}_j\cup \mathcal{H}_j}\kappa  p_j \right)
\end{equation}
and the constraint we use is for the following function to be equal to $0$, which means that the entire capacity is used. This condition could be set as an inequality (larger than or equal to $0$) in the case of streaming where the whole capacity is only used in the local minima.

\begin{equation}\label{eq:3tier_cons}
\mathcal{C} - \sum_{j\in \{1,2,3\}}\left( \sum_{i\in \mathcal{L}_j} R_ix_i  +  \sum_{i\in \mathcal{H}_j}T_j + r_j (1-T_j/R_ix_i) \right) = 0
\end{equation}
In case of file downloads, $r_j = T_j$ at optimal point (Lemma \ref{lemma:r-equals-T}), and for tier $j\in \{1,2,3\}$, we have $H_j = \{R_i: (R_i> T_j)\&(p_i == p_j)\}$ and $L_j =\{R_i:(p_i == p_j)\} - H_j$. To ensure the NLP converges, we manually pick a starting point for our solver for which the constraint in Equation \ref{eq:3tier_cons} is non-negative. 

\begin{figure}[t]
\centering
\subfigure[Initial user regrets after one round of regret minimization through the leader (ISP).]{\includegraphics[width=0.35\linewidth, angle=0]{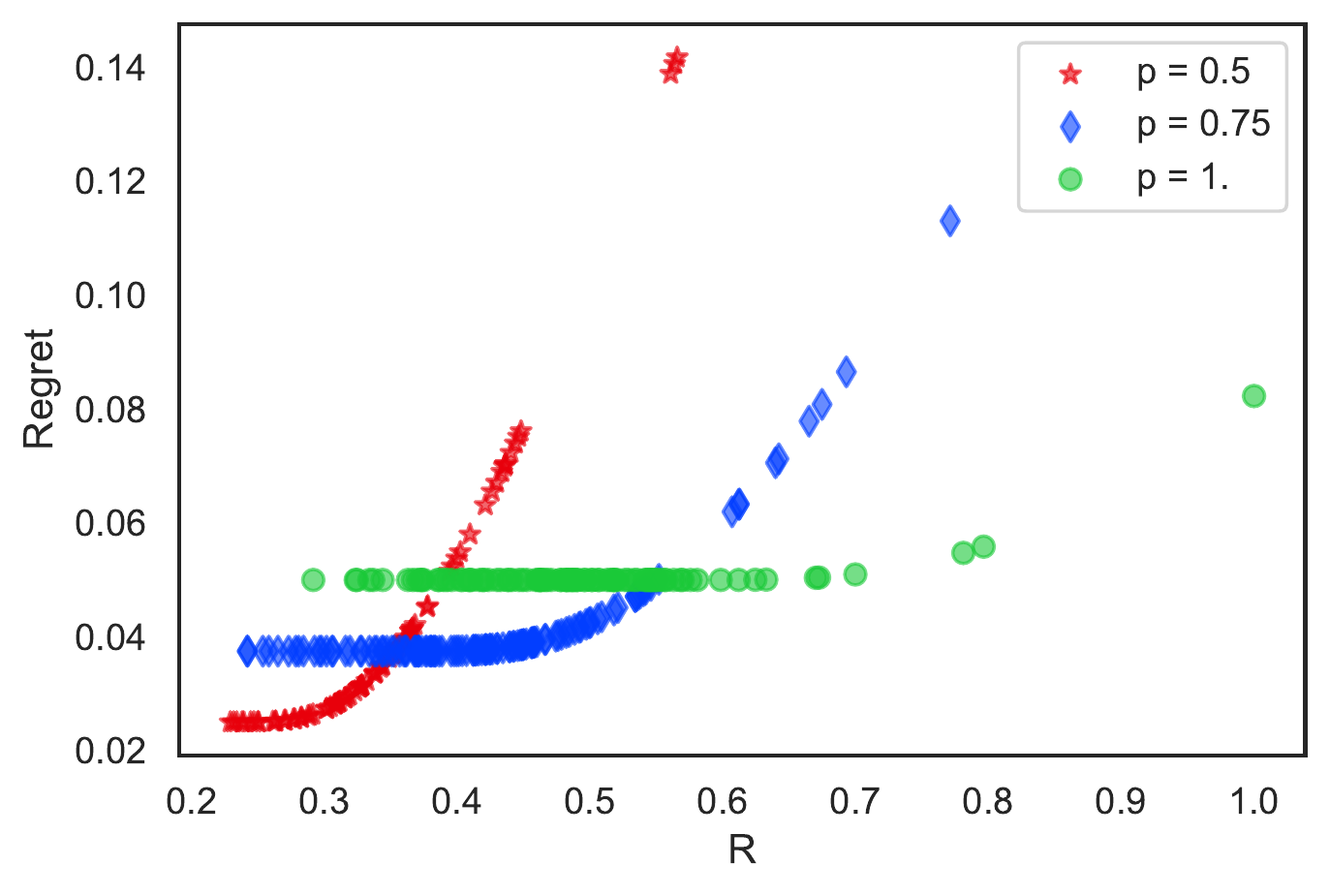}} 
\subfigure[User regrets after the leader-follower (ISP-users) game equilibrium.]{\includegraphics[width=0.35\linewidth, angle=0]{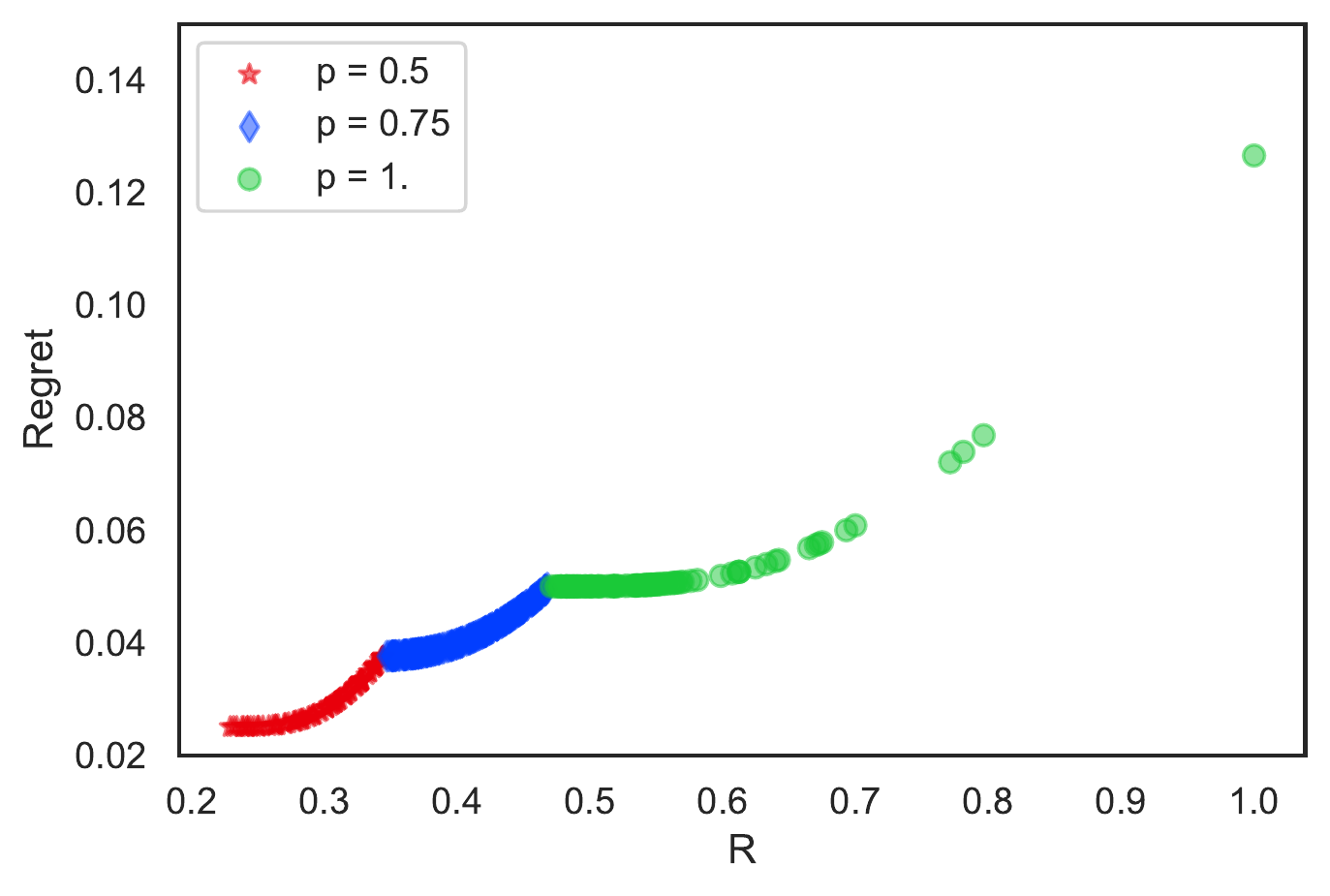}} 
\caption{The individual regrets as a function of $R$, as well as the tiers each user is choosing before (a) and after (b) the equilibrium. There are $300$ users with log-normal usage rate distribution where $\mu = 1$ and $\sigma = 0.25$. We have $\mathcal{C} = 0.95 \sum_{i\in \mathcal{N}} R_ix_i$ and $\kappa = 0.05$.}
\label{fig:3tiers}
\end{figure}

We now go through an example of the equilibria with $300$ user rates drawn from a log-normal distribution with $\mu = 1$ and $\sigma = 0.25$ and randomly assigned to tier $j$ with price $p_j$ where $p_j\in \{0.5, 0.75, 1.\}$. For the simulation of the initial user assignments, we assume low rate users have a higher probability of choosing the cheapest tier with $p_1 = 0.5$, and higher rate users move toward the more expensive tiers. To that end, the users are assigned to different tiers over a binomial trial with $2$ coin flips resulting in numbers of $0$ heads (tier $1$), $1$ head (tier $2$), and $2$ heads (tier $3$). The first $1/3$ of the users with the lowest rates have the binomial probability of $P(heads) = 0.2$, the second $1/3$ of the users have $P(heads) = 0.5$, and the remaining ones have $P(heads) = 0.8$.

In a scenario resembling the Stackelberg game \cite{von2010market}, the ISP is the leader of the game and decides the allocation among three tiers to minimize the total regret. The users are the followers who may change their tiers to minimize their individual regrets. In response to them, the ISP may again change its allocation, and so on. Figure \ref{fig:3tiers}(a) depicts the user regrets where the users are assigned to different tiers using the binomial trial explained above. After a series of leader-follower games, the system reaches an equilibrium where the user regrets and their chosen tiers are depicted in Figure \ref{fig:3tiers}(b). 

We observe that the equilibrium happens where users with the lowest rates choose the cheapest tier, and users with the highest rates choose the more expensive one. In other words, those who use more pay more, which we believe is a fair way to minimize the total regret, meet the ISP's capacity constraint, and have the users pay for their services according to their bandwidth requirements. While it is intuitive that more expensive tiers offer higher $(T_j^*,r_j^*)$, and users who have higher bandwidth demands use the more expensive tiers, our model makes it possible for the ISP to decide on the allocation $\mathcal{C}_j$ of each tier while ensuring maximum customer satisfaction.

\section{Related Work}\label{sec:related}
Some prior work consider ISP models where additional payment is required once bandwidth is exceeded instead of a reduction in rate. For instance, Wang et al. \cite{wang2017role} introduce a model where a data cap is imposed to restrain data demand from heavy users, and usage over the data cap is charged based on a per-unit fee. They analyze the impact of a data cap on a provider’s optimal pricing and find that the data cap can be set to increase the provider’s revenue under market equilibrium. Ha et al. in Tube \cite{ha2012tube} also creates a price-based feedback control loop between an ISP and its end users. On the ISP side, it computes ``time-dependent'' prices to balance the cost of congestion during peak periods with that of offering lower prices in less congested periods. 
However, most ISPs practice throttling with flat-rate pricing, even the ones who offer unlimited bandwidth \cite{Throttled}.

Some other work provide qualitative approaches on semi-informed users' decisions given ISP data cap and fees, and do not focus on quantitative modeling
. Oaks in \cite{cabletv} reports different data caps and average fees of some providers so that customers can make a \emph{semi-informed} decision, since in a given data capacity users may still get throttled unexpectedly. Chetty et al. \cite{chetty2012you} 
explore the effects of data caps on home Internet usage in urban South Africa to show users uncertainties with regards to their bandwidth usage
. Their results demonstrate how users orient and manage their Internet access under cost constraints.  

Almhana and Liu \cite{6167539} study traffic specific throttling and show the potential effect of P2P video streaming, both with and without bandwidth throttling, on the QoS experienced by subscribers. They collect some traffic data for subscribers of an ISP and show that when ISP exercises bandwidth throttling to shape traffic, the QoS can improve. 
Goth in \cite{goth2008isp} studies another aspect of throttling, where some ISPs throttle specific applications and possibly use deep packet inspection technology to throttle application-specific P2P packets. However, this practice of throttling does not ensure privacy and net neutrality regulations as different applications are treated differently \cite{gilroy2011access,wu2003network}.

Some ISPs offer zero-rating as one of their available services \cite{bayat2021zero}, where they do not charge their customers for accessing specific services, but inevitably, those services are throttled. One of the most famous ones is the ``BingeOn'' service offered by T-mobile that zero-rates the content of some partner streaming applications while throttling their rates \cite{binge1, binge2}. While these studies have not covered the best allocation practices for zero-rated services, in this work, a zero-rated service can be looked at as another tier, and the best allocation and throttling schemes can be determined for them.

\section{Conclusion and Future Work}\label{sec:conclusion}
In this work, we built a model to allocate an ISP's limited capacity to users. The ISP could have any number of tiers with different cost options, and the users could consume data on file downloads or streaming. We show how the ISP allocates its limited capacity to the users via throttling and limiting the rates of bandwidth hogs. We introduce a regret function, which is a function of bandwidth that the users lose, and the ISP's goal is to minimize its aggregate among users. Our regret function has desirable properties, including fairness, its simple optimization, and its feasibility to generalize to multiple-tier bandwidth allocation problems. While we try to identify and explain a complex phenomenon with a simplistic model for presentation purposes, we hope our work inspires more research on fairly allocating limited resources to heterogeneous users.

As future work, we plan to study a hybrid model where file downloads and streaming are considered together. We aim to obtain real-world data where we can look into the operators' market shares and thresholds in detail. Another avenue of research would be focusing on competition in the market, where we analyze user migration among ISPs. The ISP in a competitive market could also seek different alternatives to maximize its utility, such as saving up some of its capacity to generate revenue after allocating part of its monthly capacity to the users. 

\bibliographystyle{ACM-Reference-Format}
\bibliography{paper}

\section{Appendix}

\textit{Proof of Lemma \ref{lemma:min-y}}:
If there exists a user $i \in \mathcal{H}$ where $x_i = 1$ then the result holds trivially, since $x_i \le y_i \le 1$. Otherwise, we have that every throttled $i \in \mathcal{H}$ will increase $y_i > x_i$ to maintain the same overall consumption. A user $i \in \mathcal{H}$ would be satisfied with a $y_i < 1$ only if using this $y_i$ achieves its original desired consumption of $R_i x_i$. However, if all $i \in \mathcal{H}$ achieve their desired consumption, then \ref{eq:uni_cap} could be rewritten as $\sum_i R_i x_i$, which exceeded $\mathcal{C}$, necessitating throttling. 

\hfill $\blacksquare$

\textit{Proof of Theorem \ref{lemma:min_regret}}: Based on Lemma \ref{lemma:min-y}, for at least one of the users we have $y_i = 1$. Without loss of generality, we assume that $x_i = 1;\ \forall i\in \mathcal{N}$\footnote{This assumption is valid because $x_i$ is a multiplicand of $R_i$ in all of our equations, and is equivalent to using the change of variables $R_i x_i\rightarrow R_i$, or simply substituting $x_iR_i$ by $R_i$}. 

First, let's assume at least one user is getting throttled, and for exactly one user, user $N$, $y_N = 1$, and for the rest of the users, if throttled, $y_i = \frac{R_ix_i}{r} = \frac{R_i}{r}$. Therefore, the throttled users with $y_i = \frac{R_i}{r}$ are treated similarly as non-throttled users, since $ry_i = R_i$, and their regrets are zero. We have:
$$\mathcal{C} = \sum_{i = 1}^{N-1} R_i + T + r(1- \frac{T}{R_N})$$
Therefore, 
$$r = (\frac{\mathcal{C} - T -\sum_{i = 1}^{N-1} R_i}{R_N - T})R_N$$
Then since only user $N$ experiences regret, we have:
\begin{equation}\label{eq:reg_one_user}
\mathfrak{R} = (1 - \frac{\mathcal{C} - T - \sum_{i = 1}^{N-1} R_i}{R_N - T})^{\rho}(1-\frac{T}{R_N})^{\rho} = (\frac{\sum_{i = 1}^{N-1}R_i-\mathcal{C}}{R_N})^{\rho}
\end{equation}
Since the regret form of Equation \ref{eq:reg_one_user} is independent of $T$, $\rho$ does not impact $T$ minimizing the regret. 

Now, let's assume the set $\mathcal{H}$ of users get throttled, where $|\mathcal{H}| = H$. For all throttled users in $\mathcal{H}$, $y_i = 1$. The reason is because as mentioned earlier, if $y_i = \frac{R_i.x_i}{r}$, they are treated similarly as non-throttled users and we do not count them in the set $\mathcal{H}$.

Using equation \ref{eq:uni_cap}, we can compute $r$ given a constant set of $\mathcal{H}$ in an internal, i.e., between two consecutive points from $sorted(\{kickins \cup kickouts\})$, as below. 

\begin{equation} r = \frac{\mathcal{C} -\sum_{i\in \mathcal{L}}R_i - HT }{\sum_{i\in \mathcal{H}}(1-T/R_i)}
\label{eq:r}
\end{equation}

We then substitute $r$ into Equation \ref{eq:regret_agg} for $\rho = \tau$.

\begin{equation}
\begin{split}
\label{eq:regret_long}
 \mathfrak{R} =\sum_{k\in \mathcal{H}} \left( \frac{HR_k - \mathcal{C} +\sum_{i\in \mathcal{L}}R_i}{HR_k - (R_k\sum_{i\in \mathcal{H}}\frac{1}{R_i})T}\right. +\\
  \frac{\frac{\mathcal{C}-\sum_{i\in \mathcal{L}}R_i}{R_k}-R_k \sum_{i\in \mathcal{H}}\frac{1}{R_i}}{HR_k - (R_k\sum_{i\in \mathcal{H}}\frac{1}{R_i})T}\times T +\\
\left. \frac{\sum_{i\in \mathcal{H}}\frac{1}{R_i} - \frac{H}{R_k}}{HR_k - (R_k\sum_{i\in \mathcal{H}}\frac{1}{R_i})T}\times T^2 \right)^{\rho}
\end{split}
\end{equation}

Let's assume: 
$$a_k = (HR_k - \mathcal{C} +\sum_{i\in \mathcal{L}}R_i)$$
$$b_k = \frac{\mathcal{C}-\sum_{i\in \mathcal{L}}R_i}{R_k}-R_k \sum_{i\in \mathcal{H}}\frac{1}{R_i}$$
$$c_k = \sum_{i\in \mathcal{H}}\frac{1}{R_i} - \frac{H}{R_k}$$
$$d_k = MR_k,\ e_k = R_k\sum_{i\in \mathcal{H}}\frac{1}{R_i}$$

There, we have: 
\begin{equation}
\mathfrak{R}=\sum_{k\in \mathcal{H}}\left(\frac{a_k+ b_k T + c_k T^2}{d_k - e_k T}\right)^{\rho}
\end{equation}
If we compute the derivative to find $T$ minimizing the regret, we have :
\begin{equation} \label{eq:regret_parametric}
\begin{split}
\frac{d}{dT}\mathfrak{R} = & \sum_{k\in \mathcal{H}}\left[ \rho \times (\frac{a_k+ b_k T + c_k T^2}{d_k - e_k T})^{\rho-1} \times (\frac{1}{d_k - e_kT})^2\right. \\ & \left. \times\left((b_kd_k + a_k e_k) + 2c_kd_kT -c_ke_kT^2 \right) \right] 
\end{split}
\end{equation}

In order for this regret to have a root with $T$ independent of $\rho$, there should either be a root independent of $\rho$ in the term $a_k+ b_k T + c_k T^2$, or in the term $(b_kd_k + a_k e_k) + 2c_kd_kT -c_ke_kT^2$, or the terms should cancel out each other. Here, we show that the second case is true which make the Equation \ref{eq:regret_parametric} is zero. We have: 

\begin{equation} 
\begin{split}
&(\mathcal{C}-\sum_{i\in \mathcal{L}}R_i)\times(H - R_k\sum_{i =1}^{H}\frac{1}{R_i}) \\ &
- 2H(H - R_k\sum_{i\in \mathcal{H}}\frac{1}{R_i})T + \\&
(H\sum_{i\in \mathcal{H}}\frac{1}{R_i} - R_k(\sum_{i\in \mathcal{H}}\frac{1}{R_i})^2 ) T^2 = 0
\end{split}
\end{equation}
Therefore, 
\begin{equation}
(\mathcal{C}-\sum_{i\in \mathcal{L}}R_i) -2HT + (\sum_{i\in \mathcal{H}}\frac{1}{R_i})T^2 = 0
\label{eq:zero-der}  
\end{equation}

Which as a result, yields: 
\begin{equation}
    T = \frac{H \pm \sqrt{H^2 - (\mathcal{C} - \sum_{i\in \mathcal{L}}R_i)\times (\sum_{i\in \mathcal{H}}\frac{1}{R_i})} }{\sum_{i\in \mathcal{H}}\frac{1}{R_i}}
\label{eq:T_min_pre}
\end{equation}

The $T$ found from Equation \ref{eq:T_min_pre} is independent of $\rho$, therefore, the proof is complete in each interval. Furthermore, the acceptable minimum regret found from Equation \ref{eq:T_min_pre} needs to be a real number, $T^*$, where $0 < T^* < \hat{T}$. We prove that if the root found from \ref{eq:T_min_pre} is real, the larger root $T$ is always greater than $\hat{T}$. Therefore, the smaller $T$ satisfies the above condition.

In order for the root of Equation \ref{eq:T_min_pre} to be real, we need to have:

$$H^2 \geq (\mathcal{C} - \sum_{i\in \mathcal{L}}R_i)\times (\sum_{i\in \mathcal{H}}\frac{1}{R_i})$$

Given $\hat{T} = \frac{\mathcal{C}-\sum_{i\in\mathcal{L}}R_i}{\hat{H}}$, we divide the above equation by $H$:

$$H \geq \frac{(\mathcal{C} - \sum_{i\in \mathcal{L}}R_i)}{H}\times (\sum_{i\in \mathcal{H}}\frac{1}{R_i}) = \hat{T}\times (\sum_{i\in \mathcal{H}}\frac{1}{R_i})$$
or
$$ \hat{T}\leq \frac{H}{\sum_{i\in \mathcal{H}}\frac{1}{R_i}}\leq
\frac{H + \sqrt{H^2 - (\mathcal{C} - \sum_{i\in \mathcal{L}}R_i)\times (\sum_{i\in \mathcal{H}}\frac{1}{R_i})} }{\sum_{i\in \mathcal{H}}\frac{1}{R_i}} $$

Therefore, we have:
\begin{equation}
    T^* = \frac{H - \sqrt{H^2 - (\mathcal{C} - \sum_{i\in \mathcal{L}}R_i)\times (\sum_{i\in \mathcal{H}}\frac{1}{R_i})} }{\sum_{i\in \mathcal{H}}\frac{1}{R_i}}
\end{equation}

Which is independent of the exponent $\rho$.

\hfill $\blacksquare$

\textit{Proof of Lemma \ref{lemma:ind_reg}:} We first show that if the bandwidth allocation of a user is higher than that of another user before throttling ($R_i>R_j$), it will remain higher at the optimal point ($B_i>B_j$). Without loss of generality, let's assume $x_i = 1 \ \forall i \in \mathcal{N}$\footnote{This assumption is equivalent to using the change of variables $ R_i x_i\rightarrow R_i$}. Then since $x_i \leq y_i \leq 1$, we will have $y_i = 1 \ \forall i \in \mathcal{N}$. The bandwidth allocation for a throttled user $i$ is computed as:
\begin{equation}
B_i = T + r(1- \frac{T}{R_i})\label{eq:B}\end{equation}

Let's assume the users rate are sorted such that $R_1 \leq R_2\leq ... \leq R_N$. Therefore, the user $N$ with the highest rate gets throttled first as $T$ grows. When we compute the kickout points, the $r$ assigned to the highest $R_N$ will result in the lowest $T$ due to the reverse relation of $r$ and $T$. Similarly, user $1$ gets kicked in the last, i.e., gets throttled the last as $T$ grows. In the interval of $T$ where the user $N$ is the only one getting throttled, $T$ must be larger than the rate of all the other users (otherwise, they would get throttled as well.). Therefore, $B_N$ is larger than all other users' rates. If multiple users are throttled (users $N$ and $N-1$, $N-2$, ...), since $R_1 \leq R_2\leq ... \leq R_{N-1}\leq R_N$, then based on Equation \ref{eq:B}, we have $B_1 \leq B_2\leq ... \leq B_{N-1}\leq B_N$. Therefore, the user with a higher rate will always have higher bandwidth allocation. Note that if $\mathcal{C} \geq \sum_{i\in\mathcal{N}} R_ix_i$, no user will get throttled, which could be looked at as throttling with $T= \infty$, and the results intuitively hold. 

We now prove the second part of the lemma, i.e., $R_i>R_j$ entails $\mathfrak{R}_i>\mathfrak{R}_j$. Based on Equation \ref{eq:regret_agg}, the regret of an individual user $k \in \mathcal{H}$ is computed as $\mathfrak{R}_k = (1-\frac{ry_k}{R_kx_k})^{\rho} \times (1-\frac{T}{R_kx_k})^\tau$, which has a reverse relation with $R_k$.

\hfill $\blacksquare$

\textit{Proof of Lemma \ref{lemma:r-equals-T}:} For a simple proof, we substitute $r = T$ in Equation \ref{eq:r}. We have: 

$$ T= \frac{\mathcal{C} -\sum_{i\in \mathcal{L}}R_i - HT }{\sum_{i\in \mathcal{H}}(1-T/R_i)}$$
Hence,
$$ T\times \sum_{i\in \mathcal{H}}(1-T/R_i)= \mathcal{C} -\sum_{i\in \mathcal{L}}R_i - HT $$
Therefore, 
$$(\mathcal{C}-\sum_{i\in \mathcal{L}}R_i) -2HT + (\sum_{i\in \mathcal{H}}\frac{1}{R_i})T^2 = 0$$

Which is equivalent to Equation \ref{eq:zero-der}. Therefore, $T^* = r$.

\hfill $\blacksquare$

\textit{Proof of Theorem \ref{theorem:interval_reg}:} 
We know at the optimal point, $r = T$. If we show that in all intervals before the optimal point, the regret is decreasing, and in all intervals after the optimal point the regret is increasing, then the optimal point only occurs in one interval. We already have proved that in the optimal point, $T^* = r^*$ in Lemma \ref{lemma:r-equals-T}. Since $T$ and $r$ have a reverse relationship, when $T<T^*$, $r>T$ and vice versa. Therefore, all we need is to prove when $r>T$ ($T<T^*$), the derivative of the total regret is negative, and when $r<T$ ($T>T^*$), the derivative of the total regret is positive. We also use the term semi-convex to describe this property, meaning that the regret function is convex in each interval.

For the set $\mathcal{H}$ of throttled users, the aggregate regret is computed from Equation \ref{eq:regret_agg}. Therefore, when $\rho=\tau$ we have: 
$$\mathfrak{R} = \sum_{\mathcal{H}} (1-\frac{T}{R_i})^{\rho}(1-\frac{r}{R_i})^{\rho} = \sum_{\mathcal{H}} \mathfrak{R}_i^\rho$$

Hence:
\begin{equation}
\begin{split}
\frac{d}{dT}\mathfrak{R} &= \sum_{\mathcal{H}}n\mathfrak{R}_i^{\rho-1}[(-\frac{1}{R_i})(1-\frac{r}{R_i}) + (1- \frac{T}{R_i})(-\frac{1}{R_i})\frac{d}{dT}r] \\&= 
\sum_{\mathcal{H}}n\mathfrak{R}_i^{\rho-1}[\frac{r-T}{R_i^2}+(\frac{dr}{dT}+1)(\frac{T-R_i}{R_i^2})] 
\end{split}    
\end{equation}

On the other hand, from Equation \ref{eq:r}, we have:
\begin{equation}
\frac{dr}{dT} = \frac{-H[\sum_{\mathcal{H}}(1-\frac{T}{R_i})]+[\sum_{\mathcal{H}}\frac{1}{R_i}][C-\sum_{\mathcal{L}}R_i - HT]}{(\sum_{\mathcal{H}}(1-\frac{T}{R_i}))^2}
\end{equation}

We can then compute $\frac{dr}{dT}+1$ as below: 
\begin{equation}
\begin{split}
\frac{dr}{dT}+1 & =  \frac{-H[\sum_{\mathcal{H}}(1-\frac{T}{R_i})]}{(\sum_{\mathcal{H}}(1-\frac{T}{R_i}))^2} \\ 
& +\frac{[\sum_{\mathcal{H}}\frac{1}{R_i}][C-\sum_{\mathcal{L}}R_i - HT]}{(\sum_{\mathcal{H}}(1-\frac{T}{R_i}))^2} \\
& + \frac{(\sum_{\mathcal{H}}(1-\frac{T}{R_i}))^2}{(\sum_{\mathcal{H}}(1-\frac{T}{R_i}))^2}
\end{split}
\end{equation}

For simplification, let's substitute the following in the numerator of $\frac{dr}{dT}+1$:
$$(\sum_{\mathcal{H}}(1-\frac{T}{R_i}))^2 = (H-T\sum_{\mathcal{H}}\frac{1}{R_i})^2 = H^2 + T^2(\sum_{\mathcal{H}}\frac{1}{R_i})^2 - 2HT\sum_{\mathcal{H}}\frac{1}{R_i}$$

Hence, we have:
$$
\frac{dr}{dT}+1 = \sum_{\mathcal{H}}\frac{1}{R_i}(\frac{r-T}{\sum_{\mathcal{H}}(1-\frac{T}{R_k})}) $$

Again, by substituting $\frac{dr}{dT}+1$ into $\frac{d}{dT}\mathfrak{R}$, we have:
\begin{equation}
    \frac{d\mathfrak{R}}{dT} = \sum_{i\in \mathcal{H}}\rho\mathfrak{R}_i^{\rho-1} (\frac{r-T}{R_i^2})[1-(\sum_{k\in\mathcal{H}}\frac{1}{R_k})(\frac{R_i -T}{\sum_{k\in \mathcal{H}} 1- \frac{T}{R_k}})]
\end{equation}

Which for each user $i$, will be equivalent to: 

$$
    \frac{d(\mathfrak{R}_i)^{\rho}}{dT} = \rho\mathfrak{R}_i^{\rho-1} (\frac{r-T}{R_i^2})[1-(\sum_{k\in\mathcal{H}}\frac{1}{R_k})(\frac{R_i -T}{\sum_{k\in \mathcal{H}} 1- \frac{T}{R_k}})]
$$
For some users, when $r>T$, we have $\frac{d(\mathfrak{R}_i)^{\rho}}{dT} < 0$, and when $r>T$, we have $\frac{d(\mathfrak{R}_i)^{\rho}}{dT} > 0$. However, for some other users, $\frac{d(\mathfrak{R}_i)^{\rho}}{dT} > 0$ when $r>T$ and vice versa.

In the first case, when $r>T$, we have: 

\begin{equation}
\begin{split}
\frac{d(\mathfrak{R}_i)^{\rho}}{dT} < 0 \Rightarrow &  (\sum_{k\in\mathcal{H}} \frac{1}{R_k})(\frac{R_i - T}{\sum_{k\in\mathcal{H}} 1- \frac{T}{R_k}}) > 1 
\\ \Rightarrow & R_i > \frac{H}{\sum_{k\in\mathcal{H}} \frac{1}{R_k}}     
 \end{split}
\end{equation}

However, in the second case when $r>T$:
\begin{equation} \label{eq:low_usage}
R_i < \frac{H}{\sum_{k\in\mathcal{H}} \frac{1}{R_k}}     
\end{equation}

The first group of users are the ones with generally higher bandwidth consumption, and the second group are the ones with lower bandwidth consumption.

Now to move forward, we continue the proof for two users. Without loss of generality, assume $R_1 < R_2$\footnote{The equality case can conceptually merge the two users, and therefore can be looked at as a very simplified case of one user.}. We have:
$$R_1 < R_2 \Rightarrow R_1 < \frac{2}{\frac{1}{R_1}+ \frac{1}{R_2}} = \frac{H}{\sum_{k\in\mathcal{H}} \frac{1}{R_k}}$$
Which is equivalent to Equation \ref{eq:low_usage} that represents low bandwidth consumption. Similarly, it can be proved for user $2$ that it has a high bandwidth consumption. Therefore, if we have two users with different pre-throttling rates, the one with a lower rate is considered to have a low bandwidth consumption, and the other one has a high bandwidth consumption. Therefore, we have $\frac{d}{dT}\mathfrak{R}_1^{\rho} > 0$ and $\frac{d}{dT}\mathfrak{R}_2^{\rho} < 0$ for $r> T$ and vice versa. 

Our goal is to find $\rho$ for which the total regret is semi-convex. In other words, when $r>T$ ($T<T^*$), we need to have:
$$\frac{d \mathfrak{R}}{dT} = \frac{d}{dT}(\mathfrak{R}_1^{\rho}+\mathfrak{R}_2^{\rho}) <0 \Rightarrow \frac{d}{dT}\mathfrak{R}_1^{\rho} < \frac{d}{dT}\mathfrak{R}_2^\rho$$

And vice versa for $r<T$ ($T>T^*$). Let's now focus on $r>T$ (for $r<T$ the inequalities will be the opposite). For user $1$:
\begin{equation}
    \begin{split}
        \frac{d\mathfrak{R}_1^{\rho}}{dT} = &  \rho\mathfrak{R}_1^{\rho-1}(\frac{r-T}{R_1^2})\left( 1- (\frac{1}{R_1} +\frac{1}{R_2})(\frac{R_1-T}{2-\frac{T}{R_1}-\frac{T}{R_2}}) \right) \\
        = &  \rho\frac{\left((R_1-T)(R_1-r)\right)^{\rho-1}(r-T)}{R_1^{2 \rho}}\left( \frac{R_1R_2-R_1^2}{2R_1R_2-T(R_1+R_2)} \right)
    \end{split}
\end{equation}

since $R_1<R_2$, $r>T$, and $T<min(R_1, R_2)$, this derivative is positive. On the other hand, the following derivative is negative:

\begin{equation}
        \frac{d\mathfrak{R}_2^{\rho}}{dT} =  \rho \frac{\left((R_2-T)(R_2-r)\right)^{\rho-1}(r-T)}{R_2^{2 \rho}}\left( \frac{R_1R_2-R_2^2}{2R_1R_2-T(R_1+R_2)} \right)
\end{equation}

For a semi-convex function, we wish for $r>T$, we have $\frac{d\mathfrak{R}}{dT} < 0$ and for $r<T$, we have $\frac{d\mathfrak{R}}{dT} > 0$. This is similar to the sign of individual regret derivative of the high consuming users, where we must have $|\frac{d}{dT} \mathfrak{R}_2^{\rho}| > |\frac{d}{dT} \mathfrak{R}_1^{\rho}|$.  \footnote{For the case of more than two throttled users, this inequality can be generalized to: $|\sum_{\text{high consuming user $i\in \mathcal{H}$}}\frac{d}{dT} \mathfrak{R}_i^{\rho}| > |\sum_{\text{low consuming user $i\in \mathcal{H}$}}\frac{d}{dT} \mathfrak{R}_1^{\rho}|$ } 
Therefore, $|\frac{d}{dT} \mathfrak{R}_2^{\rho}| / |\frac{d}{dT} \mathfrak{R}_1^{\rho}|>1$. On the other hand, if $|\frac{d}{dT} \mathfrak{R}_2^{\rho}| / |\frac{d}{dT} \mathfrak{R}_1^{\rho}|<1$, the regret function will be semi-concave. We have: 

\begin{equation}
        |\frac{d}{dT} \mathfrak{R}_2^{\rho}| / |\frac{d}{dT} \mathfrak{R}_1^{\rho}| =
        \left(\frac{(R_2-T)(R_2 - r)}{(R_1 - T)(R_1 - r)} \right)^{\rho-1}(\frac{R_1}{R_2})^{2\rho-1}
\end{equation}

Below, we show that if $\rho< 1$, the regret function is semi-concave. Since $R_2 > R_1 > max(T, r)$, we have $\frac{R_2-T}{R_1 - T} > \frac{R_2}{R_1}$ and $\frac{R_2-r}{R_1 - r} > \frac{R_2}{R_1}$. Therefore: 

$$
        \frac{(R_2-T)(R_2 - r)}{(R_1 - T)(R_1 - r)} > (\frac{R_2}{R_1})^2 
$$
Also, since $\rho<1$:
$$
(\frac{R_2}{R_1})^2  > (\frac{R_1}{R_2})^{2+\frac{1}{\rho-1}}
$$
Hence:

$$
\frac{(R_2-T)(R_2 - r)}{(R_1 - T)(R_1 - r)} >(\frac{R_1}{R_2})^{2+\frac{1}{\rho-1}}
$$

Finally, since $ \rho < 1$: 

$$        
|\frac{d}{dT} \mathfrak{R}_2^{\rho}| / |\frac{d}{dT} \mathfrak{R}_1^{\rho}|  = \left(\frac{(R_2-T)(R_2 - r)}{(R_1 - T)(R_1 - r)} /(\frac{R_1}{R_2})^{2+\frac{1}{\rho-1}}\right)^{\rho-1} < 1 $$

Hence, if $\rho < 1$, the regret function is semi-concave. 

On the other hand, if $\rho\geq 2$, we have:

$$ (\frac{R_2}{R_1})^3 \geq (\frac{R_1}{R_2})^{2+\frac{1}{\rho-1}} $$

If we show $\frac{(R_2-T)(R_2 - r)}{(R_1 - T)(R_1 - r)} > (\frac{R_2}{R_1})^3 $, our proof is complete. In order to do so, we can substitute $r$ using Equation \ref{eq:r} for two throttled users. We have: 
$$r = \frac{\mathcal{C}-2T}{2-T(\frac{1}{R_1}+\frac{1}{R_2})}$$.

Therefore: 

$$\frac{(R_2-T)(R_2 - r)}{(R_1 - T)(R_1 - r)}= \frac{R_2-T}{R_1-T}\times \frac{2R_1R_2+T(R_1+R_2)-R_1\mathcal{C}}{2R_1R_2+T(R_1+R_2)-R_2\mathcal{C}}\times \frac{R_2}{R_1}$$

From before, we have $\frac{R_2-T}{R_1 - T} > \frac{R_2}{R_1}$. To prove this value is larger than $(\frac{R_2}{R_1})^3$, we need to show: 

$$\frac{2R_1R_2+T(R_1+R_2)-R_1\mathcal{C}}{2R_1R_2+T(R_1+R_2)-R_2\mathcal{C}}> \frac{R_2}{R_1}$$

Let's assume this statement is true. We have: 

\begin{equation}
\begin{split}
& \frac{2R_1R_2+T(R_1+R_2)-R_1\mathcal{C}}{2R_1R_2+T(R_1+R_2)-R_2\mathcal{C}}> \frac{R_2}{R_1}  \\
\Rightarrow & R_1^2(2R_2+T-\mathcal{C}) > R_2^2(2R_1+T-\mathcal{C}) \\
\Rightarrow & \frac{\mathcal{C}-T-2R_2}{R_2^2} < \frac{\mathcal{C}-T-2R_1}{R_1^2}
\end{split}
\end{equation}

Which is true since $R_1< R_2$.

Note that in case $ 1\leq \rho <2$, neither of these proofs are valid, and the system is not necessarily semi-concave nor semi-convex, and our simulations have also verified that. Therefore, in this paper, we use $\rho\geq 2$ for semi-convexity. 

\hfill $\blacksquare$

\end{document}